\documentclass{vldb}
\usepackage{graphicx}
\usepackage{times}
\usepackage{subfigure}
\usepackage{wrapfig}
\usepackage{picinpar}
\usepackage{color}
\usepackage{algorithm}
\usepackage{algorithmic}
\usepackage{url}
\usepackage{listings}
\usepackage{paralist}

\begin{document}
\title{DGFIndex for Smart Grid: Enhancing Hive with a Cost-Effective Multidimensional Range Index}
%
%
%
%
%


\numberofauthors{1} 
%
\author{
%
%
\alignauthor Yue Liu$^{1,6,7}$, Songlin Hu$^{1,6}$, Tilmann Rabl$^{2,10}, $Wantao Liu$^{1,6}$\\
             Hans-Arno Jacobsen$^{3,10}$,Kaifeng Wu$^{4,9}$, Jian Chen$^{5,8}$, Jintao Li$^{1,6}$\\
       \affaddr{$^6$Institute of Computing Technology, Chinese Academy of Sciences, China}\\
       \affaddr{$^7$University of Chinese Academy of Sciences, China}\\
       \affaddr{$^8$Zhejiang Electric Power Corporation, China}\\
       \affaddr{$^9$State Grid Electricity Science Research Institute, China}\\
       \affaddr{$^{10}$Middleware Systems Research Group University of Toronto, Canada}\\
       \email{$^1$\{liuyue01,husonglin,liuwantao,jtli\}@ict.ac.cn, $^2$tilmann.rabl@utoronto.ca}, \\
       \email{$^3$jacobsen@eecg.toronto.edu,$^4$kf-wu@sgcc.com.cn,$^5$chen\_jian@zj.sgcc.com.cn}\\
}

\maketitle

\linespread{.96}

\begin{abstract}
In Smart Grid applications, as the number of
deployed electric smart meters increases, massive amounts of
valuable meter data is generated and collected every day.
To enable reliable data collection and make business decisions
fast, high throughput storage and high-performance analysis of massive meter data become crucial for
grid companies. Considering the advantage of high efficiency, fault tolerance, and price-performance of
Hadoop and Hive systems, they are frequently deployed as  underlying platform for big data processing.
However, in real business use cases, these data analysis
applications typically involve multidimensional range queries (MDRQ)
as well as batch reading and statistics on the meter data. While Hive is high-performance at complex data batch
reading and ana\-lysis, it lacks
efficient indexing techniques for MDRQ.

In this paper, we propose DGFIndex, an index structure for Hive that
efficiently supports MDRQ for massive meter data. DGFIndex divides
the data space into cubes using the grid file technique. Unlike the
existing indexes in Hive, which stores all combinations of multiple
dimensions, DGFIndex only stores the information of cubes. This
leads to smaller index size and faster query processing.
Furthermore, with pre-computing user-defined aggregations of each
cube, DGFIndex only needs to access the boundary region for
aggregation query. Our comprehensive experiments show that DGFIndex
can save significant disk space in comparison with the existing
indexes in Hive and the query performance with DGFIndex is 2-50
times faster than existing indexes in Hive and HadoopDB for aggregation query, 2-5 times faster than both
for non-aggregation query, 2-75 times faster than scanning the whole table in different
query selectivity.
\end{abstract}

\section{Introduction}
With the development of the Smart Grid, more and more electric smart
meters are deployed. Massive amounts of meter data are sent to
centralized systems, like Smart Grid Electricity Information Collection System, at
fixed frequencies. It is challenging to store
these data and perform efficient analysis, which leads to the smart
meter big data problem.
For example, currently 17 million smart meters are deployed in the Zhejiang
Province, which will be increased to 22 million in next year.
As required by the standard of the China State Grid, each of them
will generate meter data once every 15 minutes (96 times a
day). Even only in a single table of electric quantity, there will
be 2.1 billion records needed to be stored and analyzed effectively daily.

The traditional solution in the Zhejiang province was based on a relational database management system (RDBMS). It implements its
analysis logics using SQL stored procedures, and builds many indexes
internally to improve the efficiency of selective data reading. It is
observed that global statistics on big tables lead to poor performance,
and the throughput of data writing is fairly low due to the indexes used
in the database system. With the increasing of the number of metering
devices and collection frequency, this situation becomes more dramatic
and the capacity of the current solution is reached. Since traditional
RDBMS exhibits weak scalability and unsatisfied performance on big data.
On top of that it is also surprisingly expensive for business users to deploy a commercial parallel database.
Hadoop \cite{hadoop}, an open source implementation of MapReduce \cite{Dean2008}
and GFS \cite{Ghemawat2003} allows users to run complex
analytical tasks over massive data on large commodity hardware clusters.
Thus, it also is a good choice for the Zhejiang Grid.
Furthermore, Hive \cite{thusoo2010hive,hive}, a warehouse-like tool built on top
of Hadoop that provides users with an SQL-like query
language, is also adopted, making it easier to develop and deploy big data applications.
As observed in our experiences in Zhejiang Grid,
because of the excellent scalability and powerful
analysis ability, Hive on top of Hadoop demonstrates its superiority in
term of high throughput storage, high efficient batch reading and analyzing
of big meter data. The problem is that, due to the lack of efficient multidimensional
indexing in Hive, the efficiency of MDRQ processing becomes a new challenge.

Current work on indexes on HDFS either focuses on one-di\-men\-sio\-nal indexes \cite{dittrich:hadoop++,eltabakh:eagle},
or mainly for spatial data type, such as point, rectangle or polygon \cite{eldawy2013demonstration,aji2013hadoop}.
They both can not perform effective multidimensional range query processing on non-spatial meter data.
Currently, Hive features three kinds of indexes
internally, the Compact Index, the Aggregate Index, and the Bitmap Index,
which can get relevant splits according to the predicates of a query. These existing indexes
are stored as a table in Hive containing every index dimension and an array of
locations of corresponding records. As the number of index dimensions
increases, the index table becomes very large.
It will occupy a large amounts of disk space
and result in low performance of MDRQs, as Hive first scans the
index table before processing.

With our observation of the meter data and queries in smart grid, we find that it has
some particular features mostly happen in general IoT (Internet of Things) system:
\begin{inparaenum}[(i)]
 \item Because the collected data is directly related to physical events, there is always a time stamp field in a record.
 \item Since meter data is a fact from physical space, it becomes unchanged after being verified and persisted in database.
 \item Since the change of the schema of meter data means carefully redesign of the system, and will definitely lead to complicate redeployment or at least reconfiguration of all end devices, the schema is almost static in a fairly long period of time.
 \item Since the business logic may require to add some constraints on more than one data column, many queries contain MDRQ characteristics.
 \item Most of the MDRQ queries are aggregation queries.
\end{inparaenum}

In this paper, taking advantage of the features of the meter data, we propose a distributed multidimensional index
structure named DGFIndex (Distributed Grid File Index), which uses
grid file to divide the data space into small cubes
\cite{nievergelt:grid}. With this method, we only need to store the
information of the cubes instead of every combination of multiple
index dimensions. This results in a very small index size. Moreover,
by storing the index in form of key-value pairs and cube-based
pre-computing techniques, the processing of MDRQ in Hive is highly
improved. Our contributions are three-fold:
\begin{inparaenum}[(i)]
  \item We share the experience of deploying Hadoop in Smart Grid and transforming legacy applications to Hadoop-based applications. We analyze and summarize the existing index technologies used in Hive, and point out their weakness on MDRQ processing, which implies the essential requirement for multiple dimensional index in traditional industry .
  \item We propose a distributed grid file index structure named DGFIndex, which reorganizes data on HDFS according to the splitting policy of grid file(Thus, each table
        can only create one DGFIndex).
        DGFIndex can filter unrelated splits based on predicate, and filter unrelated data segments in each split. Moreover,
        with pre-computing technique, DGFIndex only needs to read less data than query-related data. With above techniques,
        DGFIndex improves greatly the performance of processing MDRQ in Hive.
  \item We conduct extensive experiments on large scale real-world meter data workloads and TPC-H workloads.
        The results demonstrates the improved performance of processing MDRQs using DGFIndex over existing indexes in Hive and HadoopDB.
\end{inparaenum}

Compared with current indexes in Hive and other indexes on HDFS, DGFIndex's advantages are:
\begin{inparaenum}[(i)]
    \item Smaller index size can accelerate the speed of accessing index and improve the query performance.
    \item For aggregation query, DGFIndex can efficiently perform it by only scanning the boundary of query region and directly get the pre-computed value of the inner query region.
    \item By making use of the time stamp difference and setting the time stamp of collecting data as the default index dimension,
    DGFIndex does not need to update or rebuild after inserting more data, which makes sure that the writing throughput will not be influenced by existence of DGFIndex.
\end{inparaenum}

The rest of the paper is organized as follows. In Section
\ref{sec:background}, we give details of the big smart meter data
problem and introduce the existing indexes in Hive. In Section \ref{sec:system},
we will share the experience of transforming traditional legacy system
to cost effective Hadoop based system. In Section
\ref{sec:dgfindex}, we describe DGFIndex, and give details on its
architecture, the index construction, and how it is used in the MDRQ
process. Section \ref{sec:eval} discusses our comprehensive
experiment results in detail. Section \ref{sec:discuss} shows some
findings and practical experience about the existing indexes in
Hive. Section \ref{sec:related} presents related work. Finally,
Section \ref{sec:conclusion} concludes the paper with future work.

\section{Background} \label{sec:background}

In this section, we will give an overview of the \emph{Big Smart
Meter Data Problem} and introduce the Hive architecture and Hive's
indexes.

\subsection{The Big Smart Meter Data Problem}
To make electric power consumption more economic,
electric power companies and grid companies
are trying to improve the precision of their understanding
of the demand of power and the trend of power consumption for
increasing time frames. In recent years, the development and broad
adaption of smart meters makes it possible to collect meter data
multiple times every day. By analyzing these data,
electric power companies and grid companies
can get valuable information about continuous, up-to-date  power
consumption
and figure out in time important business supporting results, like line loss rate, etc.

With the number of smart meters deployed and the frequency of data
collection increasing, the amount of meter data becomes very large.
The form of meter data record is illustrated in Figure \ref{fig:record}. Each
record of meter data consists of a user id, power consumption,
collection date, positive active total electricity (PATE) with
different rates, reverse active total electricity
with different rates and various other metrics. The number of unique
user ids is tens of millions in a province of China.

\begin{figure}
\begin{center}
\resizebox{\columnwidth}{!}{%
\begin{tabular}{|c|c|c|c|c|c|}\hline
\textbf{UserId} & \textbf{PowerConsumed} & \textbf{TimeStamp} & \textbf{PATE with Rate 1} & \textbf{Other Metrics}\\\hline
24012 & 12.34 & 1332988833 & 10.45 & ...\\\hline
\end{tabular}}
\end{center}
\caption{An Example of Meter Data Record}
\label{fig:record}
\end{figure}

To get more statistical information,
analysts need to perform many ad-hoc queries on these data.
These queries have multidimensional range feature.
For example, below are some typical queries:

\begin{itemize}
  \item What was the average power consumption of user ids in the range 100 to 1000 and dates in the rangs "2013-01-01" to "2013-02-01"?
  \item How many users exist with a power consumption between 120.34 and 230.2 in the date range from "2013-01-01" to "2013-02-01"?
\end{itemize}

Additionally, many timing work flows  are executed to analysis
these meter data (stored procedures in previous RDBMS, will be described in Section \ref{sec:system}).
Many HiveQL predicates in these work flows have the same characteristics with the above ad-hoc queries.
Thus, an efficient multidimensional range index is crucial for
processing these queries in Hive.

\subsection{Hive}
Hive is a popular open-source data warehousing solution built on top
of Hadoop. Hive features HiveQL, an SQL-like declarative language.
By transforming HiveQL to a DAG (Directed Acyclic Graph) flow of
MapReduce jobs, Hive allows users to run complex analysis expressed
in HiveQL over massive data. When Hive reads the input table, it
first generates a certain number of mappers based on the size of
input table, every mapper processes a segment of the input table,
which is named a split. Then, these mappers filter data according to
the predicate of the query. Tables in Hive can be
stored in different file formats, for example, plain text format
(TextFile) and binary format (SequenceFile and RCFile
\cite{he2011rcfile}). Even though each file format can be compressed with
different compression algorithms,
Hive still has to scan the whole table without the help of index. This in turn
results in large amounts of redundant I/O, and leads to high cost of
resources and poor performance, especially
for queries with low selectivity.

Index is a powerful technique
to reduce data I/O and to improve query performance. Hive provides
an interface for developers to add new index implementations. The
purpose of an index in Hive is to reduce the number of input splits
produced by the predicate in a query. As a result, the number of mappers
will also be reduced. In the current version of Hive, there
are three kinds of indexes, the Compact Index \cite{hive:compact},
the Aggregate Index \cite{hive:aggregate}, and the Bitmap Index
\cite{hive:bitmap}. All the three types are stored as a Hive table,
and their purpose is to decrease the amount of data that needs to be read.

\begin{sloppypar}
For the Compact Index, the schema of the index table is shown in
Table \ref{tab:compact schema}. If the base table is not
partitioned, Hive uses the HiveQL statement shown in Listing
\ref{lst:index} to populate the index table. The
\lstinline$INPUT_FILE_NAME$ represents the name of input file. The
\lstinline$BLOCK_OFFSET_INSIDE_FILE$ represents the line offset in
the  \emph{TextFile} format and the \emph{SequenceFile} format and
the block offset in the \emph{RCFile} (not to be mistaken with the
block in HDFS). Compact Index stores the position information of
all combinations of multiple index dimensions of different data
files.
\end{sloppypar}

\lstset{language=SQL}
\begin{lstlisting}[caption=The Creation of a Compact Index, label=lst:index]
INSERT OVERWRITE TABLE IndexTable
SELECT <index dimension list>,
       INPUT_FILE_NAME,
       collect_set(BLOCK_OFFSET_INSIDE_FILE)
  FROM BaseTable
 GROUP BY <index dimension list>,
          INPUT_FILE_NAME
\end{lstlisting}

\begin{table}
\centering
\caption{Schema of a 3-Dimensional Compact Index }
\label{tab:compact schema}
\begin{tabular}{|c|c|}
\hline
$\textbf{Column\ Name}$ &$\textbf{Type}$\\
\hline
$index\ dimension\ 1$ &$type\ in\ base\ table$\\
\hline
$index\ dimension\ 2$ &$type\ in\ base\ table$\\
\hline
$index\ dimension\ 3$ &$type\ in\ base\ table$\\
\hline
$\_bucketname$ &$string$\\
\hline
$\_offset$ &$array<bigint>$\\
\hline
\end{tabular}\\
\end{table}

An Aggregate Index is built on the basis of Compact Index, its
purpose is to improve the processing of the \lstinline$GROUP BY$ query type. The user
can specify pre-computed aggregations when creating an Aggregate Index (for now,
only the $Count$ aggregation is supported). The schema of
an Aggregate Index' table includes additional pre-computed
aggregations at the end of every line in a Compact Index table. The
Aggregate Index uses the idea of \emph{index as data}. Using query
rewriting technique, it changes the \lstinline$GROUP BY$ query on
the base table to a scan-based query over the smaller index table.
Unfortunately, the use of Aggregate Indexes is heavily restricted:
the dimensions that are referenced in \lstinline$SELECT$,
\lstinline$WHERE$, and \lstinline$GROUP BY$ should be in the index
dimensions, and the aggregations in a query should be in the
pre-computed aggregation lists or can be derived from them.

The Bitmap Index is a powerful structure for indexing columns with a
small amount of distinct values. In the RCFile format, except
storing the offset of block, it stores the offset of every row in
the block as a bitmap. In TextFile format, every line is seen as a
block, so the offset of every row in the block is 0. Thus, Bitmap Index
only improves the query performance on RCFile format data. A
Bitmap Index changes the type of \lstinline$_offset$ in the compact
index to \lstinline$bigint$ and adds a column \lstinline$_bitmaps$
with type \lstinline$array<bigint>$.

Partition is another mechanism to improve query performance. Every
partition is a directory in HDFS. It is similar to partitioning in
a RDBMS. Partition can be seen as a coarse-grained index. The difference
of partition with above indexes is that it needs to reorganize data into different
directories.

When Hive processes query with Compact Index or Bitmap Index, it
first scans the index table and writes relevant $filename-offsets$
pairs to a temporary file. Afterwards, the method $getSplits$ in
$InputFormat$ reads the temporary file, and gets all splits from
these file names in it. Finally, $getSplits$ filters irrelevant
splits based on the offsets in temporary file.

Compact Index and Bitmap Index are used to filter unrelated splits,
and Aggregate Index is used to improve the performance of
\lstinline$GROUP BY$ queries. These three kinds of indexes have
several limitations when processing MDRQs:
\begin{enumerate}
\item When the number of distinct values in every index dimensions
is very large,
the number of records in index table will be huge. The reason is
that tables of these three types of indexes store all the combinations of each
index dimensions. This leads to excessive disk consumption and,
ultimately, a bad query performance.
\item When the records of
an index dimension that have the same value are scattered
evenly in the file (for example, every split has one record), these indexes
will not filter any splits. The reason is that they do not reorganize data
to put these records together, and their processing unit is split.
\item If the output temporary file of index is very big,
it may overflow the memory of master, because the method $getSplits$ is run in a single master
and it needs to load all information of the temporary file into memory before running MapReduce jobs.
\end{enumerate}

Partitioning in Hive is not flexible, and when creating partitions
for multiple dimensions, it will create a huge amount of
directories. This will quickly overload the NameNode. In HDFS, all
metadata about directories, files, and blocks are stored in the
NameNode's memory. The metadata of every directory occupies 150
bytes memory of NameNode \cite{cloudera}. For example, if we create partitions from
three dimensions with 100 distinct values each, 1 million
directories will be created, and 143 MB will be occupied in the
NameNode's memory, which is not including the metadata of files and
blocks. However, partition is a good complement for index, because an
index can be created on the basis of each partition.

\section{Smart Grid Electricity Information Collection System} \label{sec:system}

In this section, we will describe the data flow, system migration experience
from a RDBMS-based system to a RDBMS and Hadoop/Hive-based system of
the Electricity Consumption Information Collection System in Zhejiang Grid.

\subsection{Data Flow in Zhejiang Grid} \label{subsec:data_flow}

\begin{figure*}[t!]
\begin{center}
\includegraphics[width=0.75\textwidth,height=9cm]{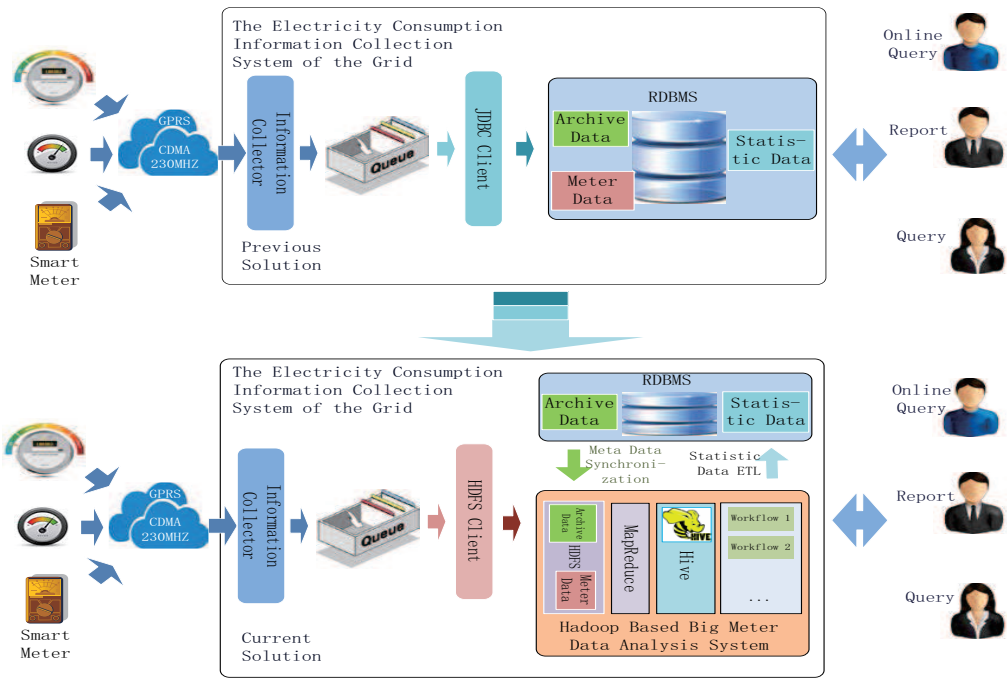}
\caption{Previous Solution and Current Solution in Zhejiang Grid}
\vspace{-10pt}
\label{fig:data_flow}
\end{center}
\end{figure*}

Figure \ref{fig:data_flow} shows the data flow abstraction in Zhejiang Grid.
The smart meters are deployed in resident houses, public
facilities and business facilities etc. The reporting frequency of
smart meter can be set. The more frequent, the more precise. The
reported meter data is transmitted by a information collector service to several
queues. The clients of RDBMS then get the data from queues and
write them into meter data tables in database.

In Zhejiang Grid, the data can be classified into \textbf{three
categories}. The \textbf{first} one is meter data which is collected
at fixed frequency. Its features are:  (1) massive amounts of data, (2)
a time stamp field in every record, (3) no changes are performed once meter data
is verified and persisted into the database, and (4) the schema of meter data is almost static.
The \textbf{second}
category is archive data which records the detailed archived information of
meter data, for example, user information of a particular smart
meter, information of power distribution areas, information
of smart meter device, etc. The archive data has different features compared
to meter data: (1) the amount is relatively small, (2) archive data is not static. The
\textbf{third} category is statistical data. The data analysis in
Zhejiang Grid consists of many off-line function modules. Each
module is in the form of a stored procedure (in the previous solution, which
will be described in Section \ref{subsec:system_migration}). Each stored procedure
contains tens of SQL statements. These stored procedure are executed at fixed
frequencies to compute, for example, data acquisition rate, power   
calculation, line loss analysis, terminal traffic statistics etc. Some SQL statements in each stored procedure join a
particular class(or several classes) of meter data with corresponding archive data
to generate statistic data and populate related tables. The
statistic data can be accessed by consumers or decision makers in the
Zhejiang Grid. Except function modules, the data analysis in the Zhejiang Grid
also includes ad-hoc queries. These queries are
dynamic compared to the function modules.

\subsection{System Migration Experience} \label{subsec:system_migration}
Based on the description of the features of data and data flow in the Zhejiang
Grid, there are mainly three requirements for the Electricity Consumption
Information Collection System in Zhejiang Grid:
\begin{enumerate}
 \item \textbf{High write throughput}. The current collected data flow needs to be written
 onto disk before the next data flow arrives.
 Otherwise, cumulative meter data will overflow the queues. Some records in current data flow may be lost.
 This is forbidden in Zhejiang Grid, because complete meter data analysis is crucial
 for power consuming monitoring and power supply adjustment. Also, the collected data
 typically is incomplete, which will
 influence the accuracy of analysis result.
 \item \textbf{High performance analysis}. High performance analysis enables more timely
 analysis report for decision makers, which makes it
 possible to adjust the power supply on demand more precisely.
 \item \textbf{Flexible scalability}. The current meter data scale has increased 30 times
 since 2008. As the collecting frequency
 and the number of deployed smart meters increase, the meter data scale will
 grow rapidly. The system should be flexibly scaled as the data scale.
\end{enumerate}

\begin{figure}[H]
\begin{center}
\includegraphics[width=0.8\columnwidth,height=4cm]{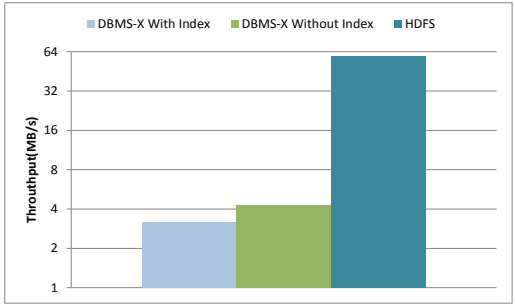}
\caption{DBMS-X vs HDFS Write Performance}
\vspace{-10pt}
\label{fig:rdbmsvshdfs}
\end{center}
\end{figure}

Figure \ref{fig:data_flow} shows the previous solution and current solution
of the Electricity Consumption Information Collection System of the Zhejiang Grid.
The upper figure shows the previous solution - an RDBMS-based storage and analysis system,
this solution mainly relies on a commercial RDBMS deployed on high-end servers.
Considering the requirements above and the meter data explosion in
Zhejiang Grid, we can easily determine that the RDBMS in the previous solution will become
the bottleneck, mainly because of three
reasons: (1) Weak scalability. RDBMSes usually depend on horizontal
sharding and vertical sharding to scale out or upgrading hardware to scale up to
improve the performance. In each scale out, the developers need to redesign
the sharding strategy and the logic in applications.
In each scale up, the system maintainer needs to buy more powerful
hardware. In both cases, each improvement will lead to huge cost of human and
financial resources. (2) Low write throughput. Figure \ref{fig:rdbmsvshdfs}
shows the write performance in real environment of Zhejiang Grid.
DBMS-X is a RDBMS from a major relational database vendor,
which is deployed on two high-end servers. The Hadoop is deployed on
13 commodity servers cluster. We can see that the write performance of DBMS-X
is much lower than HDFS. If the table in DBMS-X has an index, the write
performance will be worse. The result in Figure
\ref{fig:rdbmsvshdfs} is consistent with the findings in
\cite{pavlo2009comparison, stonebraker2010mapreduce}. (3)
Resources competition. Putting on-line transaction processes and
off-line analysis processes together in single RDBMS will aggravate the
performance. On top of that, the commercial RDBMS license is
very expensive.

The lower figure in Figure \ref{fig:data_flow} shows the current solution - a Hadoop / Hive and RDBMS-based
storage and analysis system. Hadoop / Hive
are mainly utilized for data collecting, data analysis and ad-hoc query. The RDBMS is still a
good choice for online query and CRUD operations on the archive data.
The current solution combines the advantage of off-line batch processing
and high write throughput of Hadoop ecosystem with the advantage of
powerful OLTP of RDBMS. It releases the burden of RDBMS on big data analysis,
and make Hadoop a data collecting / computing engine for the whole system. After
system migration, the burden on the RDBMS is much smaller than before.
Introducing such an cost-efficient open source
platform into Smart Grid is not costly, on the contrary, it will
consequently make it possible to reduce the "heavy armor" configuration of  the RDBMS-X,
which do not need to raise but significantly reduce the overall budget.
The IT team and the financial manager meet in their way to the business objective.
Another advantage is that open source Hadoop ecosystem can be customized
for Smart Grid, for example, adding dedicated effective multidimensional index - DGFIndex, one of
our efforts for improving the performance of Hadoop/Hive on big data analysis of Zhejiang Grid.

In the current solution,  meter data is directly written into HDFS by multiple HDFS clients.
Although, HBase\cite{hbase} as an alternative to HDFS could also enable high write throughput
and support more update operations than append as does HDFS. Also, Hive
could perform queries on HBase. However, based on our experiments using a TPC-H workload,
the query performance of Hive on HDFS is 3-4 times better than that of Hive
on HBase. To get high analysis performance, we use HDFS directly as the
storage of meter data. The archive data is stored in RDBMS, so as to
perform efficient CRUD operations by users. Furthermore, a copy of archive data
is stored in HDFS, which facilitates join operation analysis between
archive data and meter data. The two copies need to be consistent. The
analysis results (i.e., statistic data) are written into RDBMS for
online query. The SQL statements in stored procedures of the RDBMS are transformed
to corresponding HiveQL statements by our mapping tool \cite{xu2013qmapper}. The
HiveQL statements in a stored procedure are organized as work flow in Oozie \cite{oozie}. All
stored procedures, archive data synchronization, and statistic data
ETL are scheduled by the coordinator in Oozie.

Previous experiments questioned the performance of Hadoop/Hive in comparison to RDBMSs or parallel
databases \cite{pavlo2009comparison,stonebraker2010mapreduce}. However,
many improvements have be developed for Hadoop / Hive \cite{doulkeridis2013survey}
and its performance has increased significantly. One of the main reasons leading to poor performance
of Hadoop / Hive is that scan-based query processing method. To solve this problem,
adding effective index can improves Hadoop/Hive's performance dramatically \cite{jiang2010performance}.
Other Hadoop/Hive optimization for Zhejiang Grid applications we have done are presented in \cite{xu2013qmapper,hu2014dualtable}.
Furthermore, much data analysis in Zhejiang Grid is based on loops on separate regions, thus the
logic can easily be parallelized. However, in current RDBMS, the degree of parallelism
is highly limited by the configuration of the number of CPU cores, the memory capacity and the speed of disk I/O.
Higher degrees of parallelism  will need more high-end servers, which will lead to higher costs.
In a Hadoop / Hive cluster, we only need to add cheap commodity servers,
which puts less pressure on the budget. With a higher degree of parallelism, efficient indexes, and other optimizations,
Hadoop / Hive's performance can be comparable or even better than that of expensive RDBMSs.

Parallel databases could also be a candidate for data analysis in the State
Grid. But it has the same drawbacks with RDBMS, e.g., its write
throughput is much lower than
HDFS \cite{pavlo2009comparison, stonebraker2010mapreduce}.
Again, the software license cost is also high and it requires
sophisticated configuration parameter tuning and query performance optimization, and lots of maintenance efforts.

\section{DGFIndex} \label{sec:dgfindex}

In the following, we will present our novel indexing technique
DGFIndex for multidimensional range queries.

\subsection{DGFIndex Architecture}

\begin{figure}
\begin{center}
\includegraphics[width=\columnwidth]{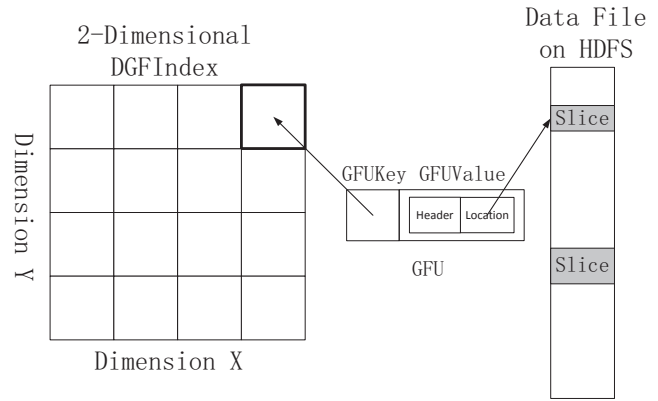}
\caption{DGFIndex Architecture}
\vspace{-10pt}
\label{fig:architecture}
\end{center}
\end{figure}

Figure \ref{fig:architecture} shows a 2-dimensional DGFIndex, in
which dimension X and Y can be any two dimensions in the record of
the meter data. DGFIndex uses a grid file to split the data space
into small units named grid file unit (\emph{GFU}), which consists
of \emph{GFUKey} and \emph{GFUValue}. \emph{GFUKey} is defined as
the left lower coordinate of each \emph{GFU} in the data space.
\emph{GFUValue} consists of the header and the location of the data
\emph{Slice} stored in HDFS. A \emph{Slice} is a segment of a file
in HDFS.
 The header in \emph{GFUValue} contains pre-computed
aggregation values of numerical dimensions, such as max, min, sum,
count, and other UDFs (need to be additive functions) supported by Hive. For example, we can
pre-compute $sum(num*price)$ of all records located in the same
\emph{GFU}.
The location in \emph{GFUValue} contains the start and end offset
of the corresponding \emph{Slice}. All
records in a \emph{Slice} belong to the same \emph{GFU}.

\begin{figure}
\begin{center}
\includegraphics[width=\columnwidth]{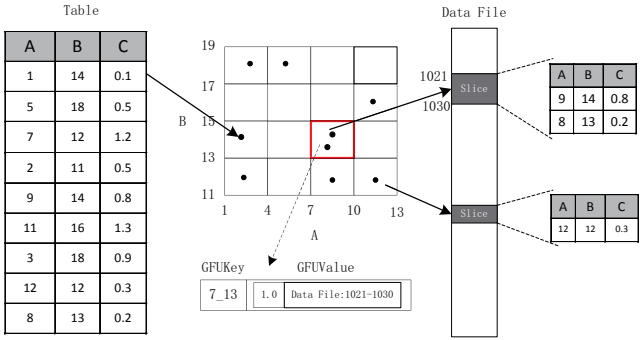}
\caption{DGFIndex Example}
\vspace{-10pt}
\label{fig:architecture_example}
\end{center}
\end{figure}

An example of a DGFIndex can be seen in Figure
\ref{fig:architecture_example}. In the example, there is a Hive
table consisting of three dimensions: A, B, and C. Suppose that the
most frequently queried HiveQL statement on this table is like the one
shown in Listing \ref{lst:HQL-example}.

\lstset{language=SQL}
\begin{lstlisting}[caption=Multidimensional Range HiveQL Query, label=lst:HQL-example]
SELECT SUM(C)
  FROM Table
 WHERE A>=5 AND A<12
   AND B>=12 AND B<16;
\end{lstlisting}

We build DGFIndex for dimensions A and B. Dimension A and B are
equally divided into intervals with granularity of 3 and 2,
respectively. Every interval is left closed-right open, e.g.,
$[1,4)$. The data space is divided into \emph{GFU}s along dimension
A and B. The records are scattered in these \emph{GFU}s. For
example, the first record $<1,14,0.1>$ is located in the region
$\{(A,B)|1 \leq A < 4, 13 \leq B < 15\}$. All records in the same
\emph{GFU} are stored in a single HDFS \emph{Slice}. Every
\emph{GFU} is a key-value pair, for example, the key-value pair of
the highlighted \emph{GFU} is as showed in Figure
\ref{fig:architecture_example}. The \emph{GFUKey} 7\_13 is the left
lower coordinate of the red one. The first part in \emph{GFUValue}
is pre-computed \lstinline$sum(C)$ of all records in the
\emph{Slice}. Users can specify any Hive supported functions and
UDFs when constructing an DGFIndex. Once a DGFIndex is deployed,
users can still add more UDFs dynamically to DGFIndex on demand. The
second part in \emph{GFUValue} is the location of the \emph{Slice}
on HDFS.

Since the index will become fairly big after many insertions, we can
utilize a distributed key-value store, such as HBase, Cassandra, or
Voldemort to improve the performance of the index access. In the
current implementation, we use HBase as the storage system for
DGFIndex.

\subsection{DGFIndex Construction}

\begin{algorithm}[H]
\caption {Map(Text line)}
\label{alg:map2}
\begin {algorithmic}[1]
\STATE $idxDValue=\textbf{getIdxDimensionList}(line)$;
\STATE $GFUKey=\phi;$
\FOR {$value$ in $idxDValue$}
\STATE $GFUKey\bigcup \textbf{standard}(value);$
\ENDFOR
\STATE $\textbf{emit} <GFUKey,line>$
\end {algorithmic}
\end {algorithm}

Before constructing a DGFIndex, one needs to specify the splitting
policy (the interval size of every index dimension) according to the
distribution of the meter data. The construction of DGFIndex is a
MapReduce job. The job reorganizes the meter data into a set of \emph{Slice}s based on the specified
splitting policy. In the meantime, it builds a
\emph{GFUKey}-\emph{GFUValue} pair for every \emph{Slice} and adds
the pair into the key-value store.
The details of the job is showed in Algorithm \ref{alg:map2} and Algorithm \ref{alg:red2}.
In the map phase, the mapper first gets all values
of index dimensions (Line 1). Then, the mapper standardizes each
value based on the splitting policy and combines these standard
values to generate \emph{GFUKey} (Lines 2-5). The "standard" method
is to find the previous coordinate in splitting policy relative to
the value on this dimension. At last, the mapper emits
$<GFUKey,line>$ to the reducer.
In the reduce phase,
the reducer first sets the start and end position of current \emph{Slice}
as the current offset of the output file of reducer and -1 respectively (Line 1-2),
and then sets the \emph{sliceSize} equal to 0, \emph{fileName} as
the current output file's name, \emph{header} as null (Line 3-5).
Second, the reducer computes all the pre-computed values and combines
them into header, the \emph{sliceSize} records the cumulative size of current \emph{Slice} (Line 6-12).
At last, the reducer computes the end position of current \emph{Slice} and GFUValue (Line 13-14),
then the reducer puts the $<GFUKey,GFUValue>$ pair into the key-value
store. In addition, the minimum and maximum standardized values in
every index dimensions are stored in the key-value store when
constructing a DGFIndex. This information is very useful when the
number of index dimension in a query is less than the number of
index dimension in the DGFIndex.

\begin {algorithm}[t]
\caption {Reduce(Text GFUKey, List$<$Text$>$ lineList)}
\label{alg:red2}
\begin {algorithmic}[1]
\STATE $start=current\ offset\ of\ ouput\ file;$
\STATE $end=-1;$
\STATE $sliceSize=0;$
\STATE $fileName=current\ ouput\ file's\ name;$
\STATE $header=NULL;$
\FOR {$line$ in $lineList$}
\STATE $preComValue=\textbf{getPreComDimensionList}(line)$;
\FOR {$value$ in $preComValue$}
\STATE $header=\textbf{combine}(header,\textbf{preCompute}(value));$
\ENDFOR
\STATE $sliceSize=sliceSize+ \textbf{sizeOf}(line);$
\ENDFOR
\STATE $end=end+sliceSize$
\STATE $GFUValue=<header,<fileName,start,end>>;$
\STATE $KVStore.\textbf{put}(GFUKey,GFUValue);$
\end {algorithmic}
\end {algorithm}

\begin{figure}
\begin{center}
\includegraphics[width=0.8\columnwidth,height=7cm]{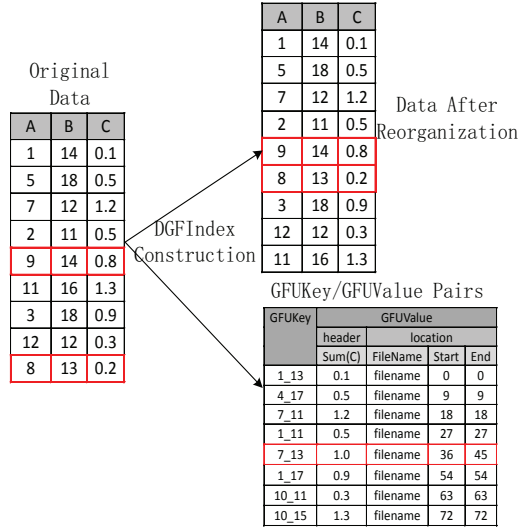}
\caption{DGFIndex Construct Example}
\vspace{-10pt}
\label{fig:construct example}
\end{center}
\end{figure}

An example of the DGFIndex construction is shown in Figure
\ref{fig:construct example}. The 5th and 9th record are located in
the same \emph{GFU}, thus, after reorganization, the two records are
stored together in a \emph{Slice}. Suppose that the size of every
record is 9 bytes. After the index construction, every
\emph{Slice} generates a $<GFUKey,GFUValue>$ pair. In
this example, we pre-compute \lstinline$sum(C)$ from every
\emph{Slice}. From the example, we can see that the maximum number
of $<GFUKey,GFUValue>$ pairs is the number of \emph{GFU}
no matter how many distinct value exist in every index dimension.The
number of records in index table is fairly small compared with the
existing indexes in Hive. For example, if we have a table containing
1000 records, we create Compact Index for 3 dimensions which have 10
distinct value respectively. There will be 1000 records in index
table, same with base table. If we create DGFIndex for these 3
dimensions with interval of 2 respectively. There are only 125
records in key-value store.

In our implementation, the syntax of constructing an DGFIndex is the
same as constructing a Compact Index in Hive except that the user
needs to specify the splitting policy and pre-computing UDF in
IDXPROPERTIES part as shown in Listing \ref{lst:createDGF}. We
specify the minimum value and interval for every index dimension.
For date type, we also need to specify the unit of interval.

\lstset{language=SQL}
\begin{lstlisting}[float=h, caption=DGFIndex Creation, label=lst:createDGF]
CREATE INDEX idx_a_b
    ON TABLE Table(A,B)
    AS 'org...dgf.DgfIndexHandler'
IDXPROPERTIES ('A'='1_3', 'B'='11_2',
               'precompute'='sum(C)');
\end{lstlisting}

There is a time stamp field in the meter data and it has been
added as a default dimension in our index. When the new meter data flow
is written into HDFS, these data first is stored in several temporary files.
After these data is verified, the time stamp dimension in DGFIndex is extended and the
DGFIndex construction process is executed on these temporary files, and the
reorganized data is written into the table directory. Thus, the data load process
is the same as the original HDFS, it does not influenced by our index.

\subsection{DGFIndex Query}
The DGFIndex query process can be divided into three steps. In the
first step, as shown in Algorithm \ref{alg:map3}, when the DGFIndex
handler receives a predicate from Hive, it first extracts the
related index value from the predicate (Line 1). If the number of
index dimension in predicate is less than the number of index
dimension in DGFIndex, the DGFIndex handler will get the minimum and
maximum standardized value of the missing index dimension from the
key-value store. Then the DGFIndex handler gets the query related
\emph{GFU}s based on the splitting policy. There are two kinds of
\emph{GFU}s, one is entirely in the query region (inner \emph{GFU})
(Line 2), another is partially in query region (boundary \emph{GFU})
(Line 3). For inner \emph{GFU}s, if the query is only an aggregation
or UDF like query, we only require the header from the key-value
store, and do not need to access data from HDFS. Thus, we can easily
get the sub result from these headers of the inner \emph{GFU}s, and
write it to a temporary file (Line 5-7). When Hive finishes computing
the boundary \emph{GFU}s, the two sub results are
combined, and returned to the user. If the query is not an aggregation
or UDF like query, we need to get all locations of the query-related \emph{Slice}s(not scanning
the index table which is different with Hive's indexes) from the key-value store,
and write them to a temporary file to help
filter unrelated splits (Line 9-12).

\begin {algorithm}
\caption {DGFIndex Query(predicate)}
\label{alg:map3}
\begin {algorithmic}[1]
\STATE $idxPred=\textbf{extract}(predicate);$
\STATE $innerKeySet=DGFIndex.\textbf{search}(idxPred);$
\STATE $boundaryKeySet=DGFIndex.\textbf{search}(idxPred);$
\STATE $queryKeySet=boundaryKeySet$
\IF {$isAggregationQuery$}
\STATE $subResult=KVStore.\textbf{getHeader}(innerKeySet);$
\STATE $\textbf{writeToTmpFile}(subResult);$
\ELSE
\STATE $queryKeySet=queryKeySet\bigcup innerKeySet;$
\ENDIF
\STATE $sliceLoc=KVStore.\textbf{getLocation}(queryKeySet);$
\STATE $\textbf{writeToTmpFile}(sliceLoc);$
\end {algorithmic}
\end {algorithm}

In the second step, as shown in Algorithm \ref{alg:red3}, the
algorithm is implemented in \lstinline$DgfInputFormat.getSplits()$.
The process is similar to the Compact Index. First, we get the set
of \emph{Slice}s from the temporary file in Algorithm \ref{alg:map3}
(Line 2). Then, we get all splits according to the name of files in
the set of \emph{Slice}s (Line 3). The splits that fully contain or
overlap with \emph{Slice}s will be chosen (Line 4-8). We prepare a
$<split,slicesInSplit>$ pair for every chosen split to filter
unrelated \emph{Slice}s in split (Line 9-12). The $sliceInSplit$ is
ordered by $start$ offset of every \emph{Slice}.

\begin {algorithm}
\caption {Split Filter}
\label{alg:red3}
\begin {algorithmic}[1]
\STATE $chosenSplit=\phi;$
\STATE $sliceSet=\textbf{readFromTmpFile}();$
\STATE $allSplit=\textbf{getSplitsFromSliceSet}($sliceSet$);$
\FOR {$split$ in $allSplit$}
\IF {$split\bigcap sliceSet \neq \phi$}
\STATE $chosenSplit=chosenSplit\bigcup split$
\ENDIF
\ENDFOR
\FOR {$split$ in $chosenSplit$}
\STATE $slicesInSplit=\textbf{getRelatedSlices}(sliceSet);$
\STATE $KVStore.put(split,slicesInSplit);$
\ENDFOR
\STATE $return\ chosenSplit;$
\end {algorithmic}
\end {algorithm}

In the third step, we implement a $RecordReader$ that can skip
unrelated \emph{Slice}s in a split. At the initialization of the
$RecordReader$, it gets a \emph{Slice} list from the key-value store
for the split it is processing. When the mapper invokes the $next$
function in the \emph{Record\-Reader}, we only need to read the records in
each \emph{Slice} and skip the margin between adjacent
\emph{Slice}s.

\begin{figure}
\begin{center}
\includegraphics[width=\columnwidth,height=6cm]{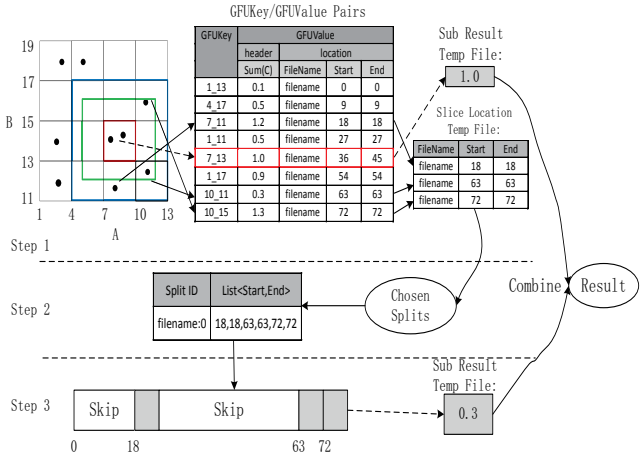}
\caption{DGFIndex Query Example}
\vspace{-10pt}
\label{fig:query example}
\end{center}
\end{figure}

The processing of the query in Listing \ref{lst:HQL-example} is
shown in Figure \ref{fig:query example}. In step 1, the query region
is $Q:\{(A,B)|5<=A< 12, 12<=B< 16\}$ (highlighted in green), the
inner region is $I:\{(A,B)|7\leq A< 10, 13\leq B< 15\}$
(highlighted in red). Because the \emph{GFU} is the smallest reading
unit for our index, the region that needs to be read is
$R:\{(A,B)|4\leq A< 13, 11\leq B< 17\}$. Consequently, the
boundary region is $R-I$. As the query is an aggregation query, we
can get a sub result from \emph{I}. Then we get the location
information of the \emph{Slice}s from the key-value store by
\emph{GFUKey}s located in boundary region. In step 2, we use
the \emph{Slice} location information to filter splits, then we
create a \emph{Slice} list for every chosen split. In our example,
we only have one split, the \emph{Slice} list means we just need to
read these regions in the split $<filename:0>$:[18-18],[63-63] and
[72-72]. Other regions can be skipped. In Step 3, we can filter
unrelated \emph{Slice}s based on the \emph{Slice} list in Step 2.

A \emph{Slice} may stretch across two splits. In this case, we divide the \emph{Slice}
into two parts: one is in previous split, another is in the adjacent split. The
two \emph{Slice}s are processed by different mappers. In our implementation, DGFIndex is transparent from the user, Hive
will automatically use a DGFIndex when processing MDRQs.

\section{Experiments and Results} \label{sec:eval}
In this section, we evaluate the DGFIndex and compare it with the
existing indexes in Hive. Furtermore, we compare
DGFIndex with HadoopDB as a comparison with parallel databases \cite{abouzeid2009hadoopdb}.
Our experiments mainly focus on three
aspects: (1) the size of index, (2) the index construction time, and
(3) the query performance.

\subsection{Cluster Setup}
We conduct experiments on a cluster of 29 virtual nodes. One node is
master for Hadoop and HBase, and the remaining 28 nodes are workers.
Each node has 8 virtual cores, 8GB memory, and 300GB disk.
All nodes run CentOS 6.5, Java 1.6.0\_45 64bit, Hadoop-1.2.1 and
HBase-0.94.13 as the key-value store. DGFIndex is implemented
based on Hive-0.10.0. Every workers in Hadoop is configured with up
to 5 mappers and 3 reducers. The replication factor is set to 2 in
HDFS. The block size is 64MB default. The mapred.task.io.sort.mb is
set 512Mb to achieve better performance. Other configurations in
Hadoop, HBase and Hive are default. For HadoopDB, we install it based
on the instructions on \cite{hadoopdb_install}. We use PostgreSQL 8.4.20
as the storage layer and above Hadoop as the computation layer.
Each experiments is run three times and we report the average result.

\subsection{DataSet and Query}
In our experiments, we use two datasets to verify the efficiency of
DGFIndex on processing of MDRQ. The first dataset is the lineitem
table from TPC-H ( 4.1 billion, about 518GB in TextFile format,
about 468GB in RCFile format, both no compression), we use it as a
general case. Another dataset is real world meter data ( 11 billion
records, about 1TB in TextFile format, about 890GB in RCFile
format, both no compression). This dataset is a kind meter data of a month.
The table comprises of 17 fields,
which contains userId, regionId( the region where the user lives),
the number of how much power consumed, and other metrics, for
example positive active total electricity with different rates etc.
These fields is not related to our queries in experiments.
The number of distinct
value in userId, regionId and time is 14 million, 11 and 30
respectively. In real world dataset, the records that have same time
are stored together, which is obey the rules of meter data. In
addition, the real world data set also contain a user's information
table which is a kind of archive data and is about 2GB. This table will be used to join with
meter data table.

We choose some queries from Zhejiang Grid for our experiments, these
queries have similar predicate with the SQLs in stored procedures. These
chosen queries mainly focus on userId, regionId, and
time. The detailed query forms are list in following parts. In each
kind of query, we change the selectivity: point query, 5\% and 12\%.
In our experiments, we suppose that there is no partitions in the
tables, if there is, we can assume that our data set is in one
partition among these partitions.

For HadoopDB, we use its GlobalHasher to partition the meter data into 28 partitions based on the userId.
Each node retrieves a 38 GB partition from HDFS. Then we use its LocalHasher
to partition the data into 38 chunks based on userId, 1GB each. All chunks are bulk-loaded into separate
databases in PostgreSQL. We create a multi-column index on the userId, regionId and time for each meterdata table.
The user table is also partitioned into 28 partitions based on userId. Each node retrieves
a 83 MB partition and puts it to all the databases of current node. Since the SMS in HadoopDB only supports
specific queries, we extend the MapReduce-based query code in HadoopDB to perform
the queries in our experiments.

\subsection{Real World Data Set}
\subsubsection{Index Size and Construction Time}
For indexes in Hive, we only compare DGFIndex with Compact Index,
since Compact index is the basis of Aggregate Index and Bitmap
Index. For now, our DGFIndex only supports TextFile table. So we use
TextFile table as the base table of DGFIndex. However, it is easy to
expend DGFIndex to support other file formats. For the Compact
Index,
RCFile-based
Compact Index will lead to smaller index table size, which will
improve the query performance. So, in our experiments, we choose
RCFile format table as the base table for the Compact Index.

In an initial experiment, we created a 3-dimensional (userId,
regionId, and time) Compact Index for the RCFile table. The size of
index table was 821GB, which is almost same with base table. As Hive
first scans index table before processing query, the 3-dimensional
Compact Index will not improve query performance. So, we only
created a 2-dimensional index (regionid and time, which have few
distinct values, 11 and 30, respectively). On the other hand, our
DGFIndex can easily handle this case. So, we create 3-dimensional
DGFIndex in following experiments. Because the number of distinct
value in $regionId$ and $time$ is small, we fix the interval size
for these: 1 and 1 day respectively. For dimension $userId$, we
change the interval size, as follow, to evaluate the influence of
different interval size on index size and query performance.
\begin{inparaenum}[(i)]
  \item Large: split dimension $userId$ equally to 100 intervals with large interval size.
  \item Medium: split dimension $userId$ equally to 1000 intervals with medium interval size.
  \item Small: split dimension $userId$ equally to 10000 intervals with small interval size.
\end{inparaenum}
What's more, we pre-compute $sum(powerConsumed)$ when building DGFIndex. This information will be used in Section \ref{sec:aggregate}.

In Table \ref{tab:size_time}, we can see that in the 3-dimensional
case, the construction of DGFIndex takes longer time than the
construction of the Compact Index, the reason is the base table needs
to be reorganized by shuffling all data to reducer via network and pre-computation CPU cost.
However, the size of DGFIndex is much smaller than
3-dimensional Compact Index , and almost equal or smaller than 2-dimensional
Compact Index. In addition, as the interval size
decreases, the number of intervals in $userId$ becomes larger, and
the number of \emph{GFU} also becomes larger, which leads to more $<GFUKey,GFUValue>$ pairs, thus bigger
DGFIndex size. In the 2-dimensional case, the size of Compact Index
is much smaller than 3-dimensional case. Because the number of
distinct value in $regionId$ and $time$ is very small. The
combinations of the two dimensions is much smaller than three
dimensions. But decreasing the number of index dimensions will decrease the accuracy of index.

\begin{table}
\centering
\caption{Index Size and Construction Time}
\label{tab:size_time}
\begin{tabular}{|c|c|c|c|c|}
\hline
$\textbf{Index}$ &$\textbf{Table}$ &$\textbf{Dimension}$ &$\textbf{Size}$ &$\textbf{Time}$\\
$\textbf{Type}$ &$\textbf{Type}$ &$\textbf{Number}$ &$\textbf{}$ &$\textbf{(s)}$\\
\hline
$Compact$ &$RCFile$ &$3$ &$821GB$ &$23350$\\
\hline
$Compact$ &$RCFile$ &$2$ &$7MB$ &$1884$\\
\hline
$DGF-L$ &$TextFile$ &$3$ &$0.94MB$ &$25816$\\
\hline
$DGF-M$ &$TextFile$ &$3$ &$3MB$ &$25632$\\
\hline
$DGF-S$ &$TextFile$ &$3$ &$13MB$ &$26027$\\
\hline
\end{tabular}\\
\end{table}

\subsubsection{Aggregation Query} \label{sec:aggregate}
\begin{figure*}[t!]
\begin{center}
\begin{minipage}[b]{0.3\textwidth}
    \centering
    \includegraphics[width=\columnwidth,height=4cm]{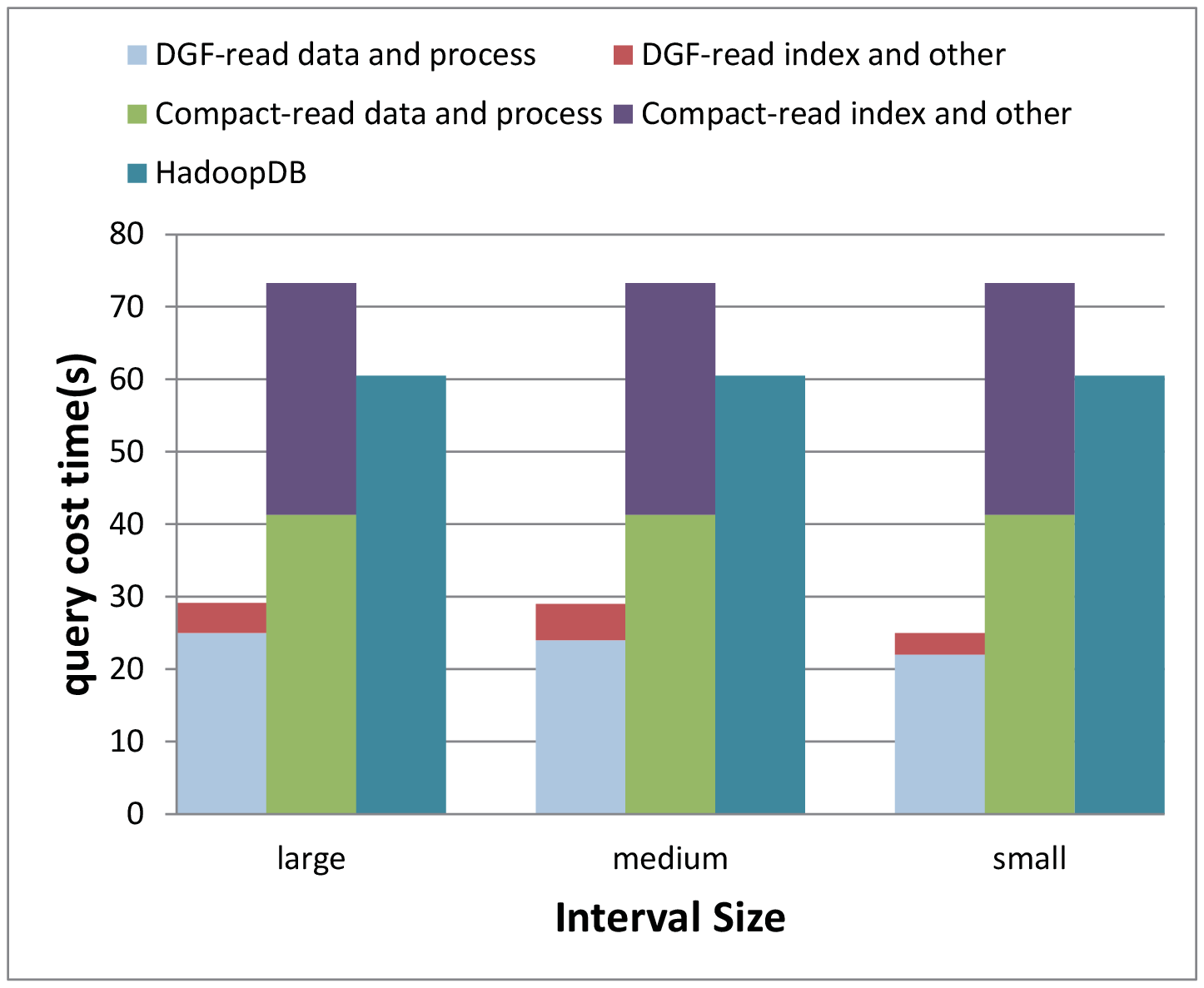}
    \caption{Aggregate Query Time For Point Query}
    \label{fig:aggregate_point_time}
\end{minipage}%
\hspace{0.1cm}
\begin{minipage}[b]{0.3\textwidth}
    \centering
    \includegraphics[width=\columnwidth,height=4cm]{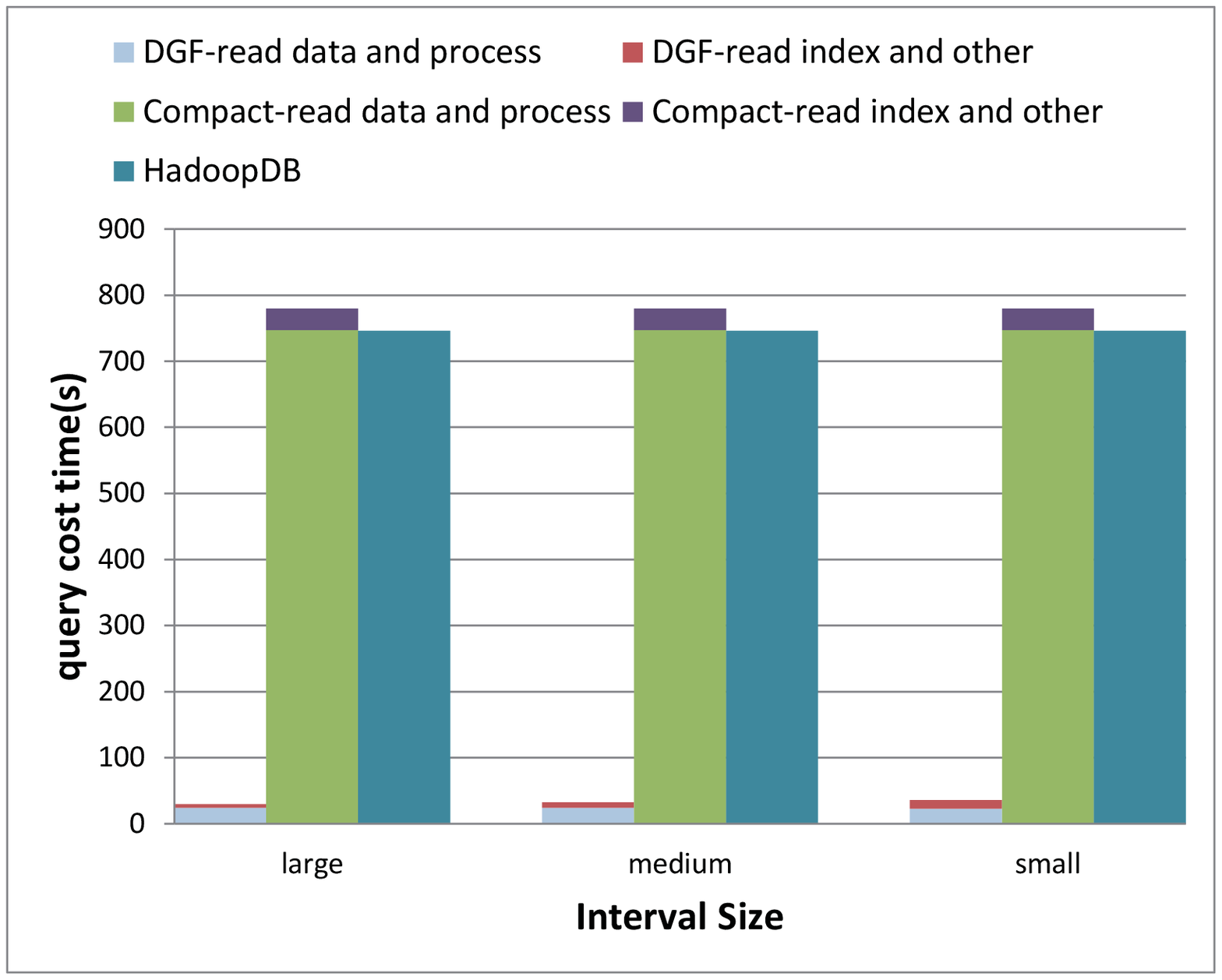}
    \caption{Aggregate Query Time For 5\% Selectivity}
    \label{fig:aggregate_5_time}
\end{minipage}%
\hspace{0.1cm}
\begin{minipage}[b]{0.3\textwidth}
    \centering
    \includegraphics[width=\columnwidth,height=4cm]{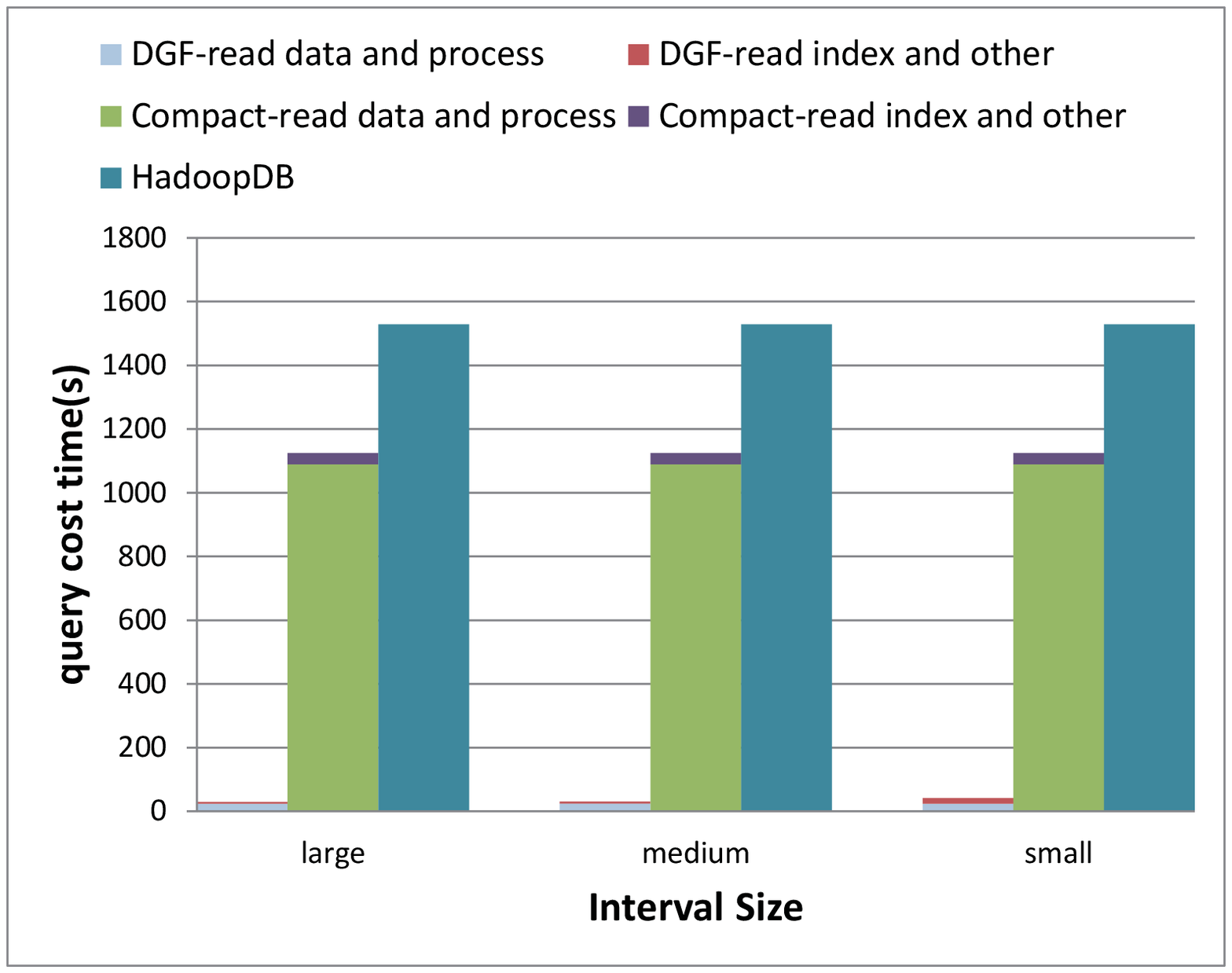}
    \caption{Aggregate Query Time For 12\% Selectivity}
    \label{fig:aggregate_12_time}
\end{minipage}

\begin{minipage}[b]{0.3\textwidth}
    \centering
    \includegraphics[width=\columnwidth,height=4cm]{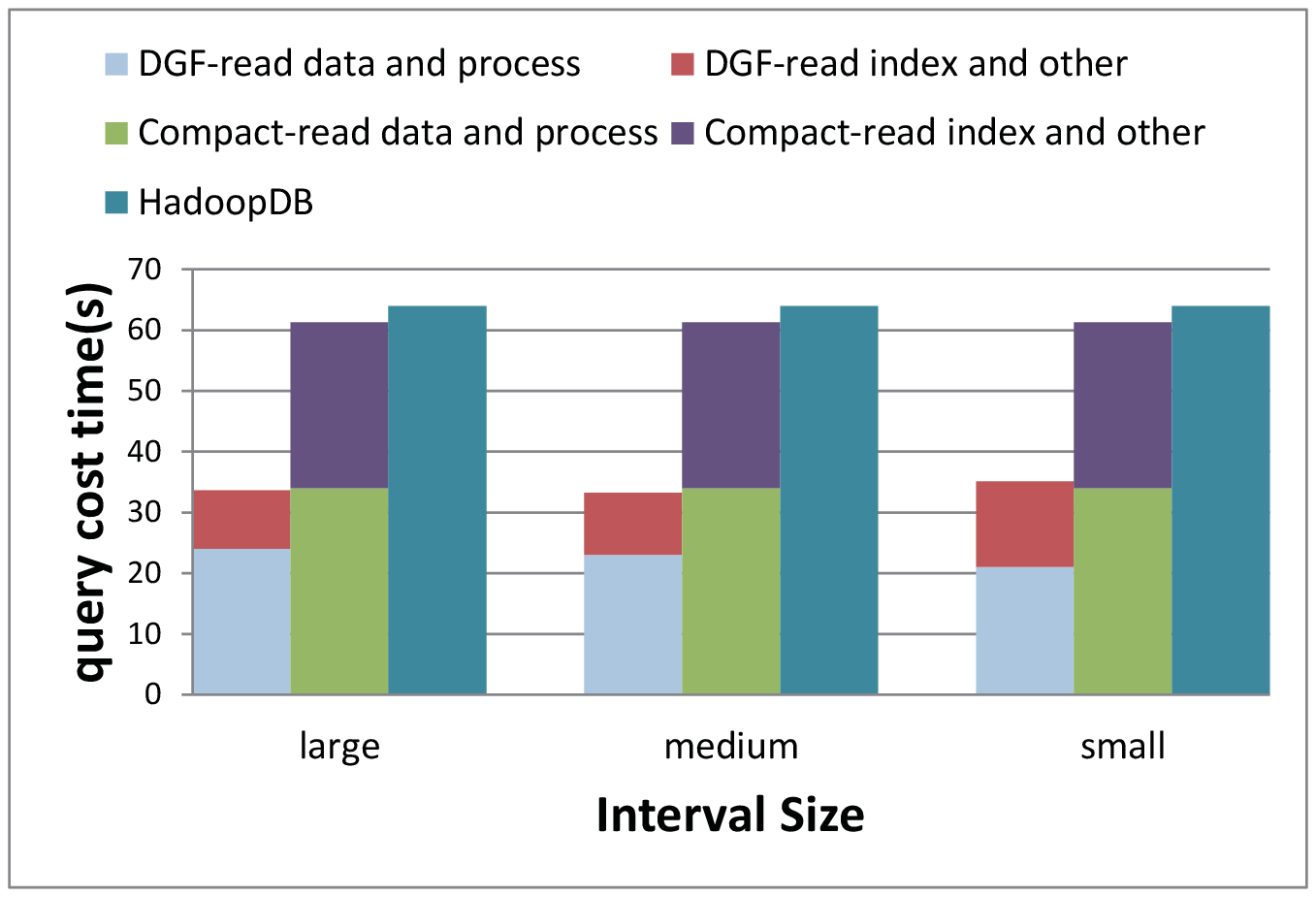}
    \caption{Group By Query Time For Point Query}
    \label{fig:groupby_point_time}
\end{minipage}%
\hspace{0.1cm}
\begin{minipage}[b]{0.3\textwidth}
    \centering
    \includegraphics[width=\columnwidth,height=4cm]{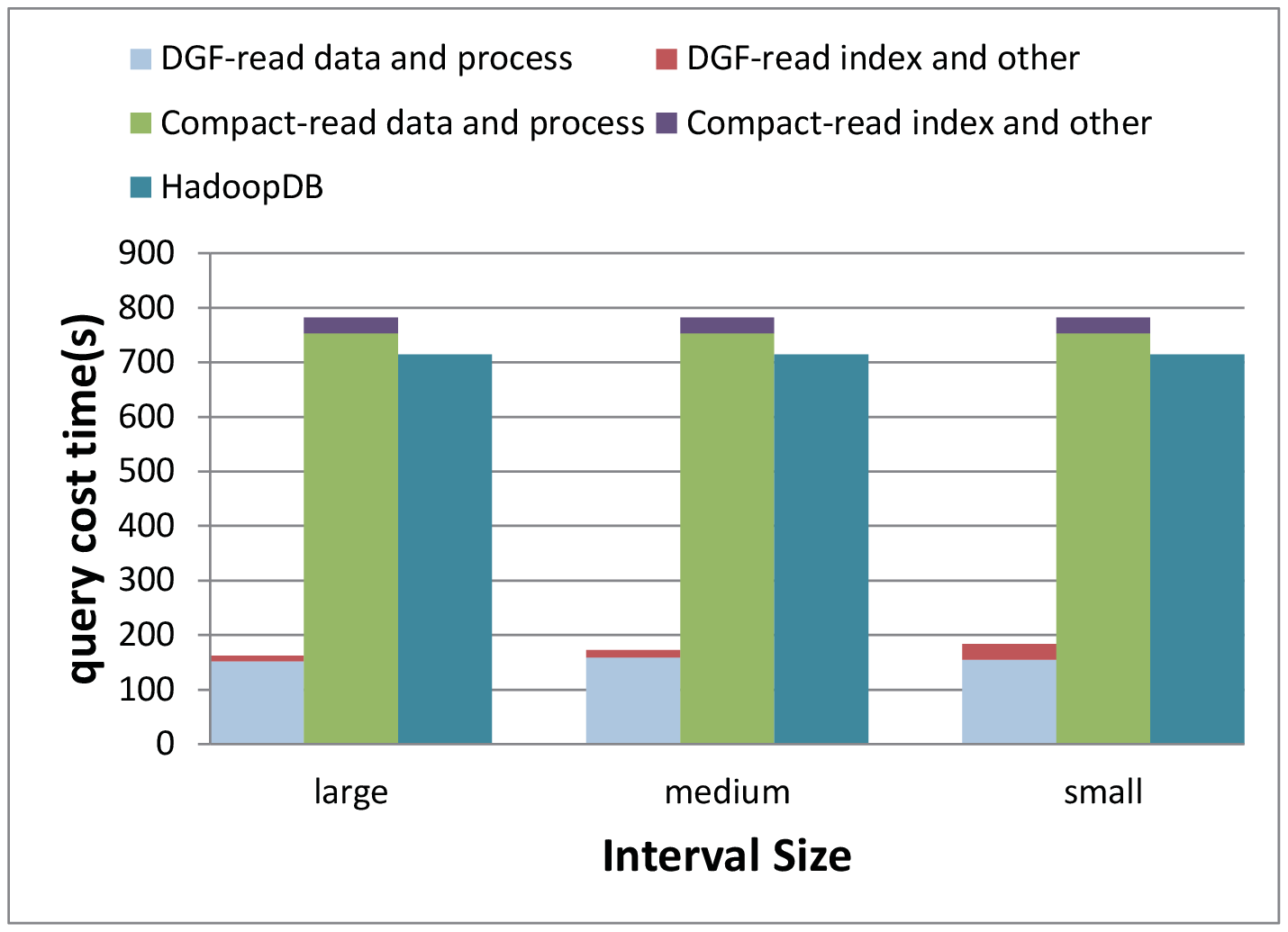}
    \caption{Group By Query Time For 5\% Selectivity}
    \label{fig:groupby_5_time}
\end{minipage}%
\hspace{0.1cm}
\begin{minipage}[b]{0.3\textwidth}
    \centering
    \includegraphics[width=\columnwidth,height=4cm]{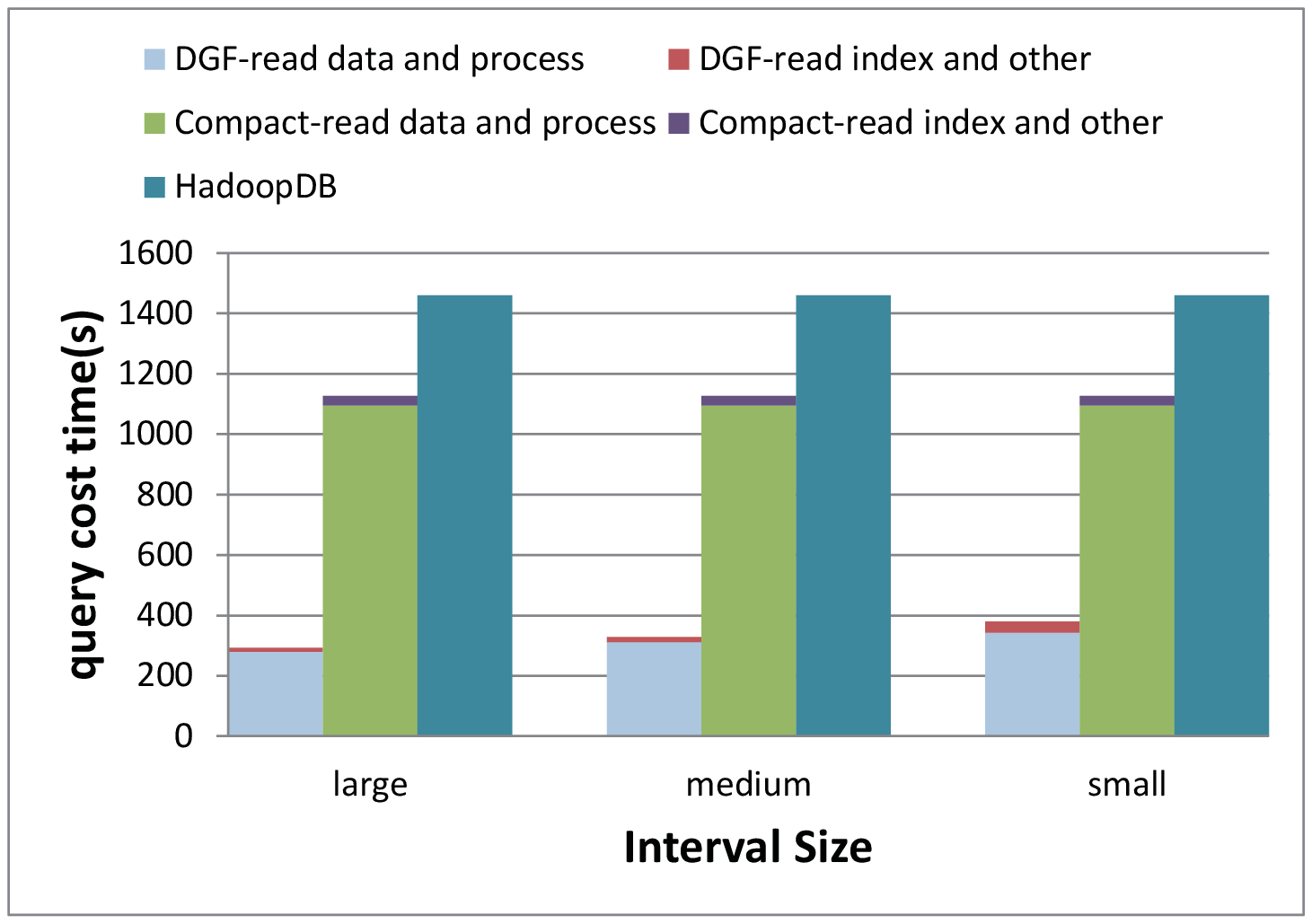}
    \caption{Group By Query Time For 12\% Selectivity}
    \label{fig:groupby_12_time}
\end{minipage}

\begin{minipage}[b]{0.3\textwidth}
    \centering
    \includegraphics[width=\columnwidth,height=4cm]{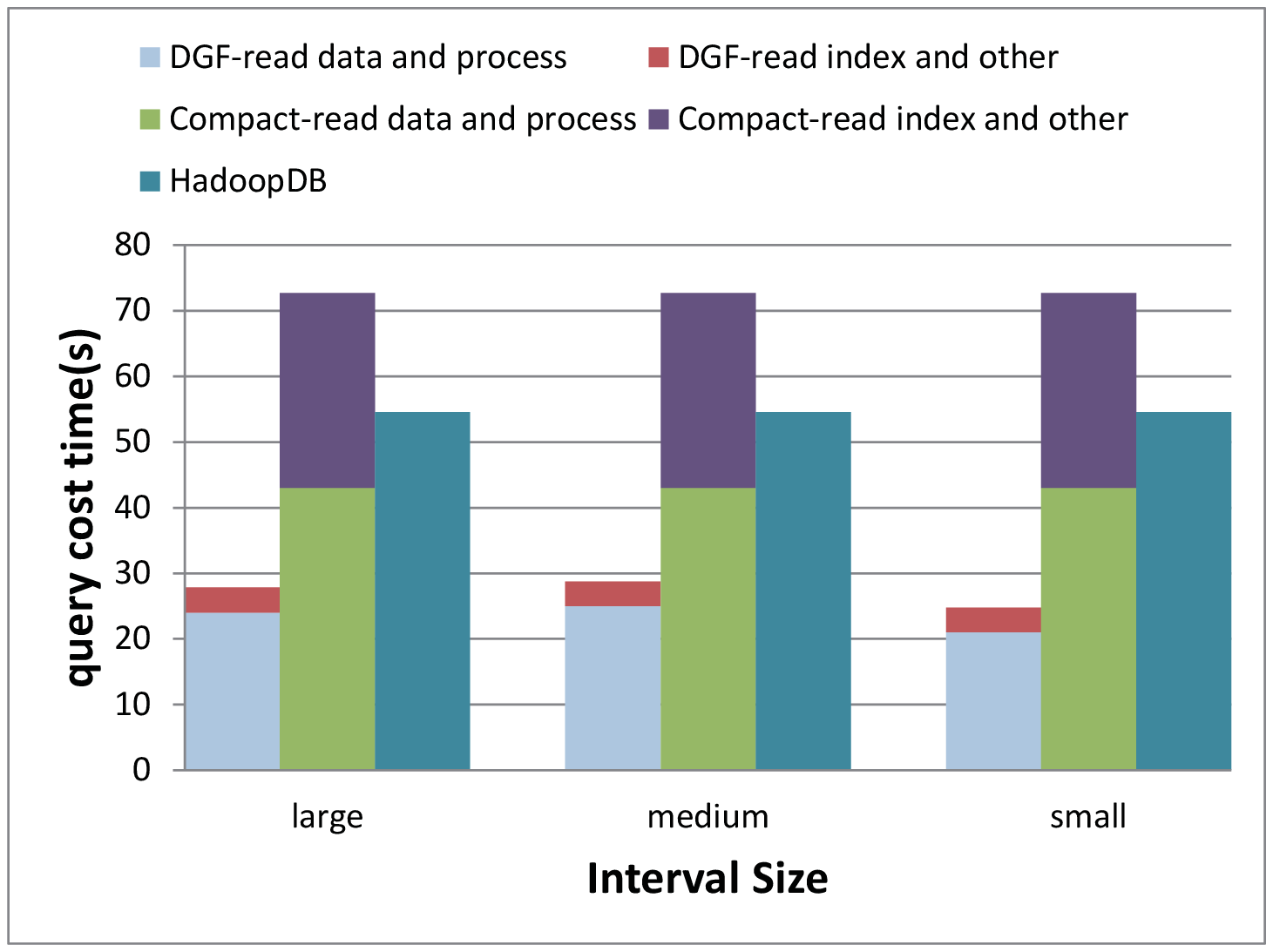}
    \caption{Join Query Time For Point Query}
    \label{fig:join_point_time}
\end{minipage}%
\hspace{0.1cm}
\begin{minipage}[b]{0.3\textwidth}
    \centering
    \includegraphics[width=\columnwidth,height=4cm]{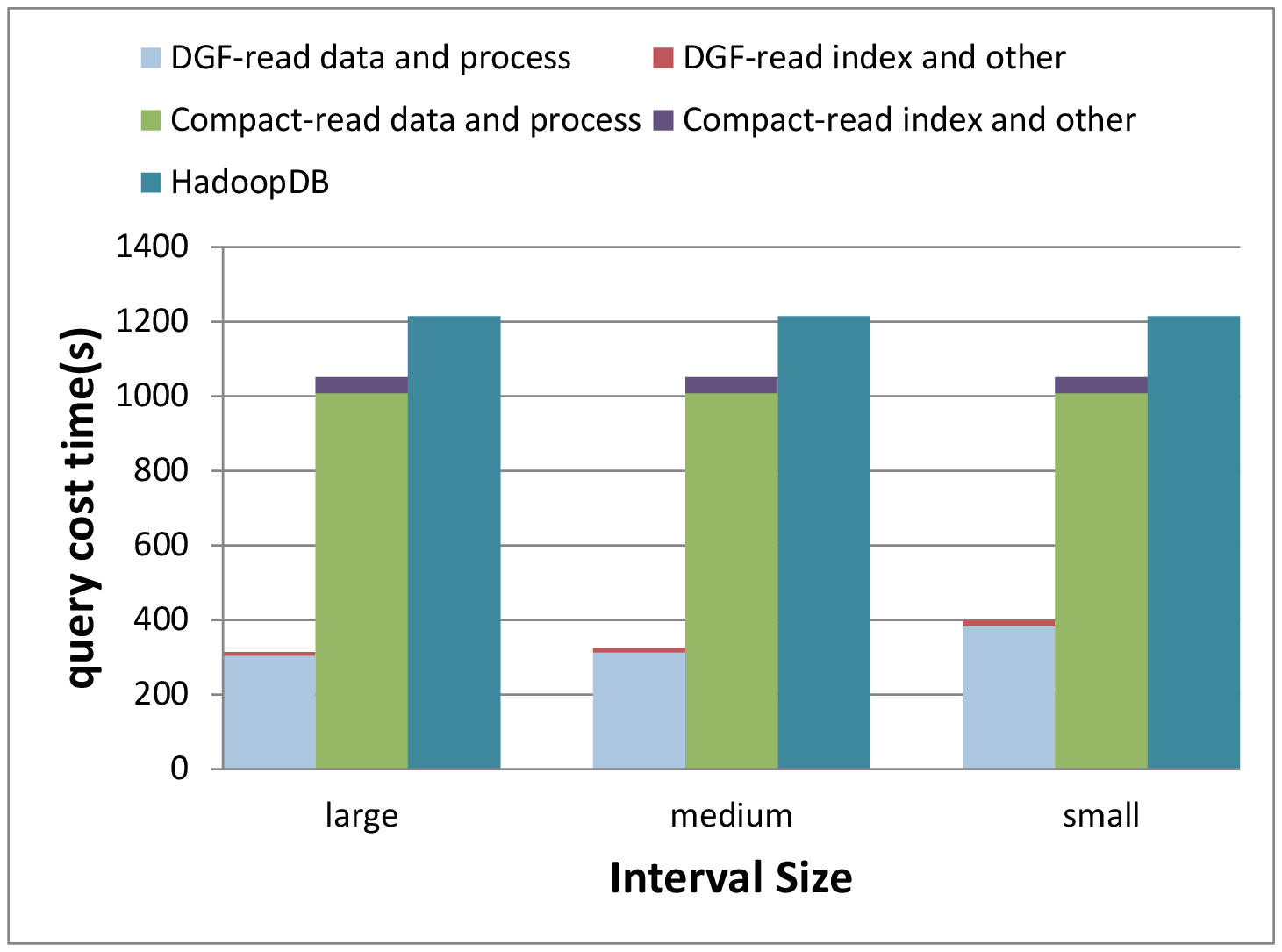}
    \caption{Join Query Time For 5\% Selectivity}
    \label{fig:join_5_time}
\end{minipage}%
\hspace{0.1cm}
\begin{minipage}[b]{0.3\textwidth}
    \centering
    \includegraphics[width=\columnwidth,height=4cm]{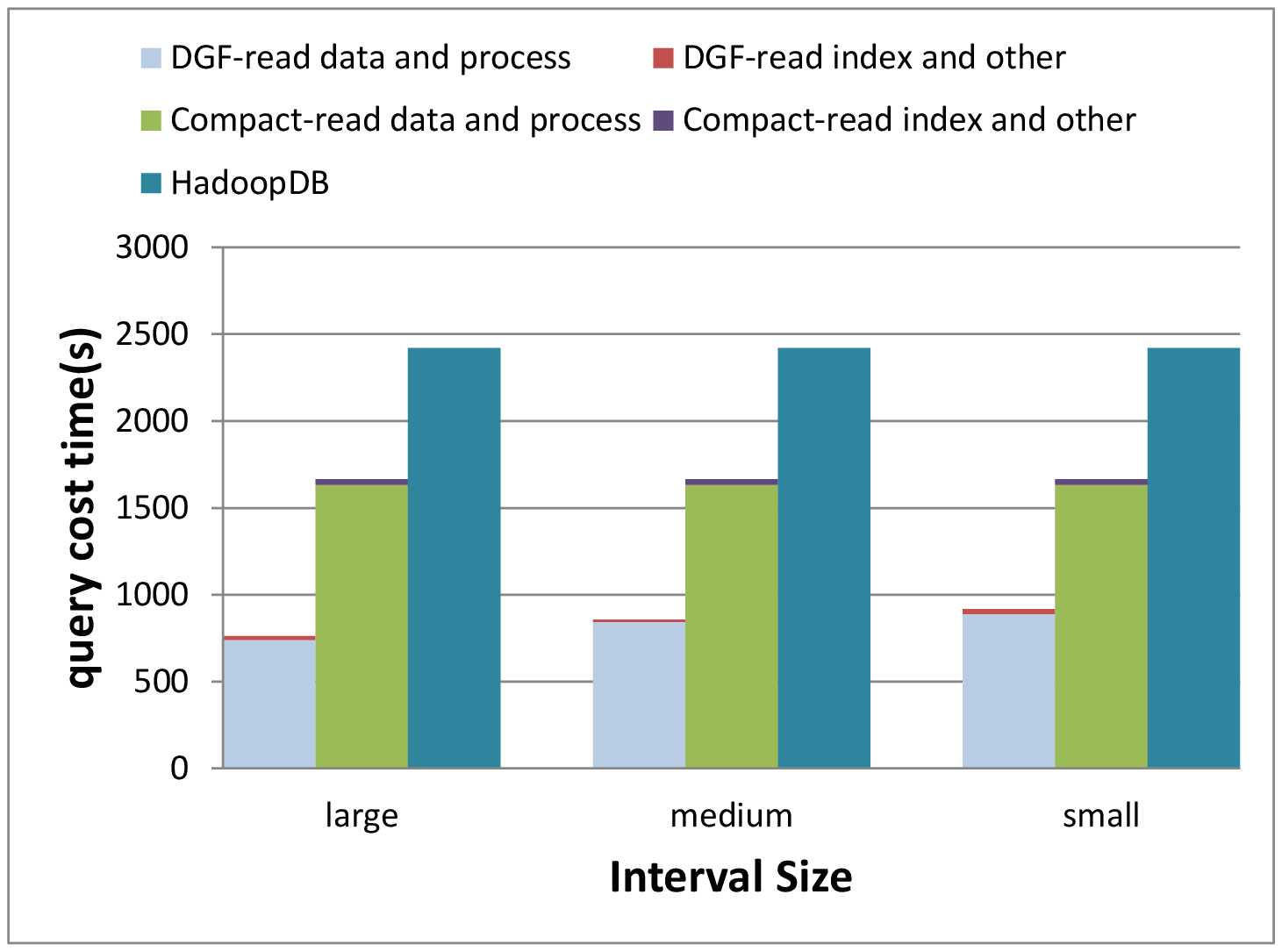}
    \caption{Join Query Time For 12\% Selectivity}
    \label{fig:join_12_time}
\end{minipage}
\vspace{-10pt}
\end{center}
\end{figure*}

In this part, we will demonstrate the efficiency of pre-computing
in DGFIndex. We choose a query like Listing
\ref{lst:aggregate_query}. Figure \ref{fig:aggregate_point_time},
\ref{fig:aggregate_5_time}, and \ref{fig:aggregate_12_time} show the
cost time of this query in different selectivities. The upper part
of the first two columns in the figures is the time of reading the index and
other time(like HiveQL parsing time and launching task time). The
lower part is the time of reading the data after filtered by
the index and processing. The third column is the cost time of HadoopDB. In HadoopDB,
the query in Listing \ref{lst:aggregate_query} is pushed into all PostgreSQL databases,
and then we use a MapReduce job to collect the results.
Table \ref{tab:aggregate_number} shows how many records
needed to read after filtered by DGFIndex and Compact Index. We do not show
the number for HadoopDB, because it is not easy to get
the number out of PostgreSQL after filtering by the index. The $accurate$ in the table
means the accurate number of records specified by predicate.

The ScanTable-based time for this kind query is about 1950s.
From the result, we can see that Compact Index
improves the performance about 26.6, 2.5, 1.7 times over scanning the whole table in different selectivity.
In large interval size case, DGFIndex improves the performance about
66.9, 65, 65.5 times over scanning the whole table. In medium interval
size case, it improves the performance 67, 59.9, 63.9 times. In small
interval size case, it improves the performance 78.1, 54.6, 46.2 times over
scaning the whole table. The number of HadoopDB is 32.2, 2.6, 1.3 respectively.
What's more, because of pre-computing, the performance of DGFIndex on
aggregation query processing nearly does not be influenced by the query selectivity.
HadoopDB has the almost same performance with Compact Index on processing aggregation query.
DGFIndex is almost 2-50 times faster than the
Compact Index and HadoopDB. We also find that HadoopDB has some performance degradation
with the selectivity increasing compared with Compact Index. The reason is that
when PostgreSQL processes multiple concurrent queries, it will lead to resources competition, and
the low batch reading performance of RDBMS is another reason.

Because we pre-compute $sum(powerConsumed)$ when constructing
DGFIndex, Hive only reads the data located in the boundary region.
From Table \ref{tab:aggregate_number}, we can see that with the decrease of interval size,
the size of $GFU$ also decrease, that is, the accuracy of DGFIndex increases, so Hive needs to read less data.
In point query case, there is no inner $GFU$, so Hive needs to read
all data located in the $GFU$. Because Compact Index
can not filter unrelated data in each splits, Hive will read the whole split, which
leads to more data reading.

\lstset{language=SQL}
\begin{lstlisting}[caption=Aggregation Query, label=lst:aggregate_query]
SELECT sum(powerConsumed)
FROM meterdata
WHERE regionId>r1 and regionId<r2
    and userId>u1 and userId<u2
    and time>t1 and time<t2;
\end{lstlisting}

\begin{table}
\centering
\caption{Records Number for Aggregation Query}
\label{tab:aggregate_number}
\small
\begin{tabular}{|c|c|c|c|}
\hline
$\textbf{Index Type}$ &$\textbf{Point}$ &$\textbf{5\%}$ &$\textbf{12\%}$\\
\hline
$Compact$ &$169,395,953$ &$4,756,501,768$ &$6,586,886,752$\\
\hline
$DGF-L$ &$4,347,200$ &$67,678$ &$100,386$\\
\hline
$DGF-M$ &$4,258,358$ &$20,280$ &$31,215$\\
\hline
$DGF-S$ &$2,291,718$ &$16,122$ &$23,712$\\
\hline
$Accurate$ &$26$ &$569,186,384$ &$1,354,351,336$\\
\hline
\end{tabular}\\
\end{table}

\begin{table}
\centering
\caption{Records Number for Group By Query}
\small
\label{tab:groupby_number}
\begin{tabular}{|c|c|c|c|c|}
\hline
$\textbf{Index Type}$ &$\textbf{Point}$ &$\textbf{5\%}$ &$\textbf{12\%}$\\
\hline
$Compact$ &$169,395,953$ &$4,756,501,768$ &$6,586,886,752$\\
\hline
$DGF-L$ &$4,347,200$ &$681,321,681$ &$1,433,931,728$\\
\hline
$DGF-M$ &$4,258,358$ &$641,128,331$ &$1,401,070,456$\\
\hline
$DGF-S$ &$2,291,718$ &$572,231,864$ &$1,367,754,156$\\
\hline
$Accurate$ &$26$ &$569,186,384$ &$1,354,351,336$\\
\hline
\end{tabular}\\
\end{table}
\subsubsection{GroupBy Query and Join Query}
In this part, we evaluate the performance of DGFIndex on processing
non-aggregation query, this means that DGFIndex can not use
pre-computed information. We use a \lstinline$Group By$ query like
Listing \ref{lst:group_query} and a \lstinline$Join$ query like
Listing \ref{lst:join_query}. For HadoopDB, we extend its aggregation task code and join task code to perfom the two queries.
The cost time of
\lstinline$Group By$ query is shown in Figure
\ref{fig:groupby_point_time},\ref{fig:groupby_5_time},
\ref{fig:groupby_12_time} and the cost time of
\lstinline$Join$ query is shown in Figure
\ref{fig:join_point_time}, \ref{fig:join_5_time}, and
\ref{fig:join_12_time}. Table \ref{tab:groupby_number} shows the
number of records needed to read of \lstinline$Group By$ query and \lstinline$Join$ query after they are filtered by the
index. The number is same for both query, since their predicate is the same.

In \lstinline$Group By$ query case, the ScanTable-based time is about 1900s. In
\lstinline$Join$ query case, the time is about 1930s. The
Compact Index improves performance 1.2-31 times over scanning the
whole table in both cases. HadoopDB improves performance 0.8-35.3 times over scanning the
whole table. On other hand, the number of DGFIndex is 2.1-75.8
times. From the result, we can see that DGFIndex is about 2-5 times faster
than Compact Index and HadoopDB on different selectivities. From Figure \ref{fig:groupby_5_time},\ref{fig:groupby_12_time},
we can see that the time of reading index becomes longer with the decreasing of interval
size. Because when the interval size becomes smaller,
more $GFU$s will be located in query region. The index handler needs to get
more GFUValue from HBase. However, storing index in key-value store decreases
index reading time. From Figure \ref{fig:groupby_12_time} and \ref{fig:join_12_time}, we can see that
Compact Index and HadoopDB's performance is almost equal or worse than ScanTable-based style when
processing high selectivity query. The reason for Compact Index is the inaccuracy of two dimensional index
leads to reading almost all splits. The reason for HadoopDB is resources competition and low batch reading speed.
However, for DGFIndex, since it can accurately read related data, thus it can maintain
effectiveness for high selectivity query.

As shown in Table \ref{tab:groupby_number}, as the interval sizes increase,
more data is located in one $GFU$. So a DGFIndex with the large
interval size needs to read more data than a DGFIndex with small
interval size. Which leads to some performance degradation, especially for
high selectivity query. However, since DGFIdex can filter unrelated slices
when reading each split, the amount of data read by DGFIndex is much
smaller than in the case of Compact Index, which improves Hive's performance dramatically.

\lstset{language=SQL}
\begin{lstlisting}[caption=Aggregation Query, label=lst:group_query]
SELECT time,sum(powerConsumed)
FROM meterdata
WHERE regionId>r1 and regionId<r2
    and userId>u1 and userId<u2
    and time>t1 and time<t2
GROUP BY time;
\end{lstlisting}

\lstset{language=SQL}
\begin{lstlisting}[caption=Join Query, label=lst:join_query]
INSERT OVERWRITE DIRECORY '/tmp/result'
SELECT t2.userName,t1.powerConsumed
FROM meterdata t1
JOIN userInfo t2
ON t1.userId=t2.userId
WHERE t1.regionId>r1 AND t1.regionId<r2
  AND t1.userId>u1 AND t1.userId<u2
  AND t1.time>t1 AND t1.time<t2;
\end{lstlisting}

\subsubsection{Partial Specified Query}

\begin{figure}[t]
\begin{center}
\vspace{-10pt}
\includegraphics[width=0.8\columnwidth,height=4cm]{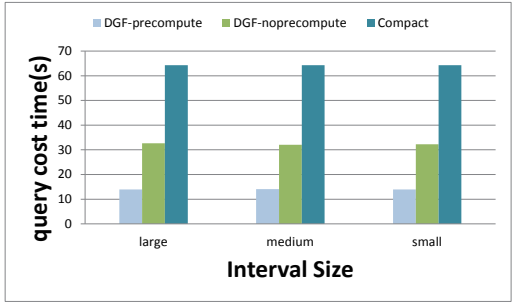}
\caption{Partial Query Time}
\vspace{-10pt}
\label{fig:partital_query}
\end{center}
\end{figure}

In practice, the number of dimensions in the predicate clause may be
more or less than the indexed dimensions. If the number is more than
indexed dimensions, DGFIndex just uses the indexed dimensions in the
predicate to filter unrelated data; If the number is less than
indexed dimensions, DGFIndex gets the minimum and maximum value of
the missing dimension from HBase to complement the predicate. This
part evaluates the second case. Because the number of $userId$
values is the largest among the three index dimensions. We delete
the $userId$ range condition from predicate, and choose a query as
shown in Listing \ref{lst:part_query}. The result is shown in Figure
\ref{fig:partital_query}. From the result we can see that DGFIndex
is 2-4.6 times faster than Compact Index.

\lstset{language=SQL}
\begin{lstlisting}[caption=Partital Query, label=lst:part_query]
SELECT SUM(powerConsumed)
FROM meterdata
WHERE regionId=11
  AND time='2012-12-30';
\end{lstlisting}

\subsection{TPC-H Data Set}

\begin{figure}[t]
\begin{center}
\includegraphics[width=0.8\columnwidth,height=4cm]{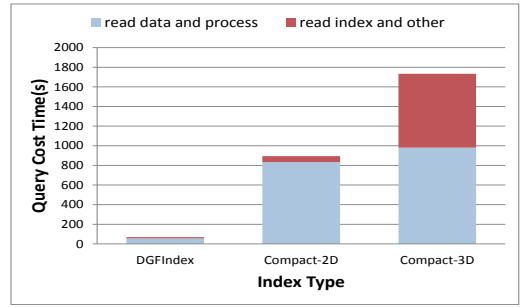}
\caption{TPC-H Workload Query Time}
\vspace{-10pt}
\label{fig:tpch_query}
\end{center}
\end{figure}

In this part, we want to demonstrate the efficiency of DGFIndex for
general case, not only for meter data. In this experiment, we use
the Q6 in TPC-H as our query. We create 2-dimensional(l\_discount
and l\_quantity, which have few distinct value) and 3-dimensional
Compact Index for lineitem table. For DGFIndex, we set the interval
size of l\_discount, l\_quantity and l\_shipdate to 0.01, 1.0 and
100 days respectively. The index size and constructing time are
showed in Table \ref{tab:size_time_tpch}. The query performance is
showed in Figure \ref{fig:tpch_query}. Table
\ref{tab:aggregate_number_tpch} shows how many records needed to
read after filtered by index.

The ScanTable-based time for this query is 632s. In this case,
2-dimensional and 3-dimensioinal Compact Index both are slower
than scanning the whole table. Because Compact Index does not filter
any split. DGFIndex is 25 times faster than Compact Index. The index
size and the number of data needed read of DGFindex is much smaller
than Compact Index. From the result, we can see that the performance
of Compact Index is much worse than real world data set. The
difference between real world data set and TPC-H data set is that
the real world data set actually is sorted by time, however, in
TPC-H data set, the records are evenly scattered in data files.
Because Compact Index does not reorganize data, it can not improve
query performance on this kind of data.

\vspace{-1em}

\begin{table}[H]
\centering
\caption{Index Size and Construction Time for TPC-H}
\label{tab:size_time_tpch}
\begin{tabular}{|c|c|c|c|c|}
\hline
$\textbf{Index}$ &$\textbf{Table}$ &$\textbf{Dimension}$ &$\textbf{Size}$ &$\textbf{Time}$\\
$\textbf{Type}$ &$\textbf{Type}$ &$\textbf{Number}$ &$\textbf{}$ &$\textbf{(s)}$\\
\hline
$Compact$ &$RCFile$ &$3$ &$189GB$ &$7367$\\
\hline
$Compact$ &$RCFile$ &$2$ &$637MB$ &$991$\\
\hline
$DGFIndex$ &$TextFile$ &$3$ &$4.3MB$ &$10997$\\
\hline
\end{tabular}\\
\end{table}

\vspace{-2em}

\begin{table}[H]
\centering
\caption{Records Number for TPC-H Workload}
\label{tab:aggregate_number_tpch}
\begin{tabular}{|c|c|}
\hline
$\textbf{Index Type}$ &$\textbf{Record Number}$\\
\hline
$Whole\ Table$ &$4,095,002,340$ \\
\hline
$Compact-3$ &$4,095,002,340$ \\
\hline
$Compact-2$ &$4,095,002,340$ \\
\hline
$DGFIndex$ &$85,430,966$\\
\hline
$Accurate$ &$77,955,077$\\
\hline
\end{tabular}\\
\end{table}

\vspace{-1em}

\section{Experience About Hive Index} \label{sec:discuss}
In this section, we will report some findings and some practical
experience about performance improvement of Hive at the aspect of
index. For now, the existing indexes in Hive are not practical and
hard to use, as there is limited documentation and usage examples.
Without reading the source code of Hive to get more information
about the indexes, they are not usable. The Compact Index is the
basis of other indexes. The number of records it stores is decided
by the number of combinations of indexed dimensions and the number
of data files. The performance of Compact Index is dependent of the
size of the index table and the distribution of the values of
indexed dimensions. If the index dimensions have few distinct values
and the data file of the table is sorted by the indexed dimension, a
good query performance improvement can be achieved using Compact
Index. On the other hand, if the indexed dimension has many distinct
values and they are evenly distributed in the data files, no
performance improvement can be achieved with Compact Index. On the
contrary, the performance will be worse than scanning the whole
table. The idea of the technique of the Aggregate Index is good, but
in practice, there are very few use case that can meet its
restrictions.

In industry, the most practical method to improve query performance
in Hive currently is partition. Partition reorganizes data
into different directories based on partition dimension.
The best way to improve Hive performance is combining partition with Compact Index.
However, the user need to make sure that the index dimensions do not have too many
distinct value and are not evenly distributed.

\section{Related Work} \label{sec:related}

As mentioned above, the existing indexes in Hive, Compact Index,
Bitmap Index and Aggregate Index, are closely related to our
DGFIndex. However, they are not well fitted to process multidimensional
range queries. \cite{abouzeid2009hadoopdb} combines MapReduce and RDBMS, and uses
the indexes in RDBMS to filter unrelated data in every worker.
But the data loading speed is lower than HDFS, and its multiple databases storage mode easily
leads to serious resource competition, especially for high selectivity query.

\begin{sloppypar}
Current various parallel databases have been applied for many
companies, such as Greenplum, Teradata, Aster Data, Netezza, Datallegro,
Dataupia, Vertica, ParAccel, Neoview, DB2(via the Database Partitioning Feature),
and Oracle(via Exadata). But these systems have
low data loading performance and the software license is very expensive.
What's more, they usually need complex configuration parameters tuning
and lots of maintenance efforts\cite{pavlo2009comparison}. Our objective
in this paper is to provide a scalable and cost effective solution for
the big meter data problem in Zhejiang Grid, the solution's performance can be comparable
or even better than parallel database, but with much low budget.
\end{sloppypar}

In the context of index on HDFS, \cite{jiang2010performance}
proposes a kind of one-dimensional range index for sorted file on
HDFS. The sorted file is divided into some equal-size pages. It
creates one index entry which comprises of the start, end value and
offset for every page. \cite{dittrich:hadoop++} proposes two kinds
of indexes, Trojan index and Trojan join index, to filter unrelated
splits and improve the performance of join tasks respectively. In
Trojan index, it stores the first key, last key and records number
for every split. \cite{eltabakh:eagle} stores the range information
for every numerical and date field dimensions in a split, and
creates a inverted index for the string type dimension. Since these
values are rarely changing, it create a materialized view to store
them in a separate file. \cite{lin2011full} creates LZO block level
full-text inverted index for data on HDFS to accelerate selection
operations that contains free-text filters. Both,
\cite{jiang2010performance} and \cite{dittrich:hadoop++} mainly
focus on one-dimensional index and they need to sort data file based
on index dimension. The primary purpose of
\cite{jiang2010performance}, \cite{dittrich:hadoop++}, and
\cite{eltabakh:eagle} is filtering unrelated splits. They can not
filter unrelated records in a split. \cite{lin2011full} can not
process multidimensional range query. In contrast, DGFIndex can
process multidimensional query efficiently, and does not need to
sort data files based on index dimensions. It only puts these
records in the same \emph{GFU} together in a file. Moreover,
DGFIndex can filter unrelated \emph{Slice}s in  a split.

In the context of spatial database on Hadoop, Spatial Hadoop\cite{eldawy2013demonstration}
proposes a two level multidimensional index on Hadoop. It first partitions
data using Grid File, R-Tree or R+-Tree into equal-size block(64MB), second creates local
index for each block, the local index is stored as a file on HDFS, third it create global
index for all block, the global index is stored in master's memory. Spatial Hadoop is mainly for
spatial data types, such as Point, Rectangle and Polygon.
Hadoop-Gis\cite{aji2013hadoop} also proposes a two level spatial index on Hadoop. It splits
data with grid partition into small size tiles, which is much smaller than 64MB. Then a global index
is created to stored the MBRs of these tiles, it is stored in the memory of master.
The local index of each tile is created on demand based on the query type. Both indexes
are applied in spatial applications, which is much different with our index, our DGFIndex is
for improving the traditional applications. Another difference is that our DGFIndex is organized
as one level index, which is simpler and easier to maintenance than the both above.  
The last difference is that we combine pre-aggregated technique with Grid File, which makes
aggregation query processing more effective.

\section{Conclusion and Future Work} \label{sec:conclusion}
\begin{sloppypar}
In this paper, we share the system migration experience from traditional RDBMS
to Hadoop based system. To improve Hadoop/Hive's performance on smart meter big
data analysis, we propose a multidimensional range index named
DGFIndex. By dividing the data space into some $GFU$s, DGFIndex only
stores the information of $GFU$ rather than the combinations of
index dimensions. This method reduces the index size dramatically.
With $GFU$-based pre-computing, DGFIndex enhances greatly the
multidimensional aggregation processing ability of Hive. DGFIndex   
first filters splits with $Slice$ location information, then skips
unrelated $Slice$ in each split. By doing this, DGFIndex reduces
greatly the number of data need to read and improve the performance
of Hive. Our experiments on real world data set and TPC-H data set
demonstrate that DGFIndex not only improve the performance of Hive
on meter data, but also is applicable to general data set. In future work, we will work on an algorithm to find the best
splitting policy for DGFIndex based on the distribution of the meter
data and the query history. The optimal placement of $Slice$s will also be
our next step research problem.
\end{sloppypar}

\textbf{Acknowledgements}: This work is supported by the National Natural Science Foundation of China under Grant No.61070027, 61020106002, 61161160566.
This work is also supported by Technology Project of the State Grid Corporation of China under Grant No.SG [2012] 815.

\begin{scriptsize}
\bibliographystyle{abbrv}
\bibliography{sigproc-sp}
\end{scriptsize}

\end{document}